\documentclass[nonacm,11pt]{acmart}

\usepackage{amsmath}%, amsthm}
\usepackage{subcaption}
\usepackage{url}
\usepackage{multirow}
\newcommand{\best}[1]{\color[HTML]{036400} \bf #1}
\newcommand{\worst}[1]{\color[HTML]{FE0000}\bf #1}

\definecolor{mygreen}{HTML}{3E833E}
\definecolor{myorange}{HTML}{ff3300}
\definecolor{myrgb}{rgb}{0, 0.125, 0.376}

\newcommand{\ignore}[1]{}
\newcommand{\Oh}{\mathcal O}
\newcommand{\A}{\Sigma}
\newcommand{\R}{\mathbf{R}}
\newcommand{\RsupMN}{\R^{n,m}}

\newcommand{\Cset}{{\mathcal C}}
\newcommand{\Ru}{{\mathcal R}}
\newcommand{\Xx}{{\mathcal I}}
\newcommand{\sy}[2]{\langle{#1,\!#2}\rangle}
\newcommand{\sz}[1]{\$}  % {Z}
\newcommand{\Val}{\mathop{\mathsf{eval}}\nolimits}
\newcommand{\Exp}{\mathop{\mathsf{exp}}}
\newcommand{\Reach}{\mathop{\mathsf{rows}}}
\newcommand{\Sum}{\mathop{\mathsf{sum}}\nolimits}

\newcommand{\gzip}{\mbox{\sf gzip}\xspace}
\newcommand{\xz}{\mbox{\sf xz}\xspace}
\newcommand{\reint}{\mbox{\sf re\_32}\xspace}
\newcommand{\reiv}{\mbox{\sf re\_iv}\xspace}
\newcommand{\reansiv}{\mbox{\sf re\_ans}\xspace}
\newcommand{\csrv}{\mbox{\sf csrv}\xspace}

\newcommand{\CSM}{\mbox{\sf CSM}\xspace}
\newcommand{\CSRV}{\mbox{$( S,V )$}\xspace}
\newcommand{\GRAMM}{\mbox{$( \Cset, \Ru,V)$}\xspace}%algorithms 

\newcommand{\susy     }{{\em Susy\/}\xspace}
\newcommand{\higgs    }{{\em Higgs\/}\xspace}
\newcommand{\covtype  }{{\em Covtype\/}\xspace}
\newcommand{\census   }{{\em Census\/}\xspace}
\newcommand{\airlinexx}{{\em Airline78\/}\xspace}
\newcommand{\optical  }{{\em Optical\/}\xspace}
\newcommand{\mnistxm  }{{\em Mnist2m\/}\xspace}

\newcommand{\MWM}{{\sf MWM}\xspace}
\newcommand{\LKH}{{\sf LKH}\xspace}
\newcommand{\PathCover}{{\sf PathCover}\xspace}
\newcommand{\PathCoverP}{{\sf PathCover+}\xspace}

\usepackage[ruled,vlined]{algorithm2e}

\setcopyright{iw3c2w3}
\copyrightyear{2022}

\begin{document}
\title{Improving Matrix-vector Multiplication via Lossless Grammar-Compressed Matrices}\thanks{This paper is published under the Creative Commons Attribution-NonCommercial-NoDerivs~4.0 International
(CC-BY-NC-ND~4.0) license. Authors reserve their rights to
disseminate the work on their personal and corporate Web sites with
the appropriate attribution.}

%%
%% The "author" command and its associated commands are used to define the authors and their affiliations.
\author{Paolo Ferragina}
\affiliation{%
  \institution{Department of Computer Science, University of Pisa}
  %\streetaddress{L.go B.~Pontecorvo, 3}
  \city{Pisa}
  \country{Italy}
  \postcode{56127}
}
%\email{surname.name@unipi.it}

\author{Travis Gagie}
%\orcid{0000-0002-1825-0097}
\affiliation{%
  \institution{Faculty of Computer Science, Dalhousie University}
  %\streetaddress{6050 University Avenue}
  \city{Halifax}
  \country{Canada}
  \postcode{NS B3H 4R2}
}
%\email{travis.gagie@dal.ca}

\author{Dominik K\"{o}ppl}
%\orcid{0000-0001-5109-3700}
\affiliation{%
  \institution{M\&D Data Science Center, Tokyo Medical and Dental University}
  %\streetaddress{2-chōme-3-10 Kanda Surugadai}
  \city{Chiyoda City, Tōkyō-to}
  \country{Japan}
  %\postcode{101-0062}
}
%\email{koeppl.dsc@tmd.ac.jp}

\author{Giovanni Manzini}
\affiliation{%
  \institution{Department of Computer Science, University of Pisa}
  %\streetaddress{L.go B.~Pontecorvo, 3}
  %\city{Pisa}
  \country{Italy}
  %\postcode{56127}
}

\author{Gonzalo Navarro}
\affiliation{%
  \institution{Millennium Institute for Foundational Research on Data}
  \institution{Department of Computer Science, University of Chile}
  \country{Chile}
}
%\email{gnavarro@dcc.uchile.cl}

\author{Manuel Striani}
\affiliation{%
  \institution{Department of Sciences and Technological Innovation,}
  \institution{University of Piemonte Orientale, Italy}
%  \country{Alessandria, Italy}
}
%\email{manuel.striani@uniupo.it}

\author{Francesco Tosoni}
\affiliation{%
  \institution{Department of Computer Science, University of Pisa}
  %\streetaddress{L.go B.~Pontecorvo, 3}
  %\city{Pisa}
  \country{Italy}
  %\postcode{56127}
}
%\email{francesco.tosoni@phd.unipi.it}

\begin{abstract}
As nowadays Machine Learning (ML) techniques are generating huge data collections, the problem of how to efficiently engineer their storage and operations is becoming of paramount importance. In this article we propose a new lossless compression scheme for real-valued matrices which achieves efficient performance in terms of compression ratio and time for linear-algebra operations.  Experiments show that, as a compressor, our tool is clearly superior to \gzip and it is usually within 20\% of \xz in terms of compression ratio. In addition, our compressed format supports matrix-vector multiplications in time and space proportional to the size of the compressed representation, unlike \gzip and \xz that require the full decompression of the compressed matrix. To our knowledge our lossless compressor is the first one achieving time and space complexities which match the theoretical limit expressed by the $k$-th order statistical entropy of the input.

To achieve further time/space reductions, we propose column-reordering algorithms hinging on a novel column-similarity score. Our experiments on various data sets of ML matrices show that, with a modest preprocessing time, our column reordering can yield a further reduction of up to 16\% in the peak memory usage during matrix-vector multiplication. 

Finally, we compare our proposal against the state-of-the-art Compressed Linear Algebra (CLA) approach showing that ours runs always at least twice faster (in a multi-thread setting) and achieves better compressed space occupancy for most of the tested data sets. This experimentally confirms the provably effective theoretical bounds we show for our compressed-matrix approach.

\end{abstract}

\authorsaddresses{}

\maketitle
\renewcommand{\shortauthors}{P. Ferragina, T. Gagie, D. K\:oppl, G. Manzini, G. Navarro, M. Striani, and F. Tosoni}

\section{Introduction}
Matrix operations have always been important in scientific computing and engineering, and they have become even more so with the widespread adoption of ML and deep learning tools. 
Very large matrices do not just present scalability challenges for their storage: they also consume a large amount of bandwidth resources in server-to-client transmissions, as well as in CPU/GPU-memory communications. Hence matrix compression appears as an attractive choice. Common simple heuristics for shrinking ML models are generally based on {\em lossy} compression, like low and ultra-low precision storage, sparsification (i.e., reduction of the number of non-zero values), and quantisation (i.e., reduction of the value domain). Unfortunately,  lossy compression schemes often impair the ML model accuracy in a data- and algorithm-specific manner, thus requiring an attentive and manual application.

For this reason {\em lossless} compression represents a better alternative for achieving ``automated'' space savings. It is data-independent and does not require any {\em a priori} knowledge about the input data. In addition, if some problem domain is not sensitive to the use of a particular lossy technique, we can apply  lossy compression followed by the lossless one, therefore getting the best of both worlds.
Unfortunately, traditional one-dimensional lossless compression techniques such as Huffman, Lempel-Ziv, bzip, Run-Length Encoding (RLE) often perform poorly on matrices, in that they are not able to unfold the (sometimes hidden) dependencies or redundancies between rows and columns. Moreover, they usually require the full matrix decompression in order to perform the needed linear algebra operations, which means that the space reduction is only achieved in the storage or transmission phases, but not in the more critical computation phase.

Recently, some authors~\cite{elgohary2019compressed,elgohary19matrix,francisco2018exploiting} have proposed new lossless compression schemes for matrices which not only save space, but also manage to speed up linear algebra operations, and matrix multiplication in particular. These results apply mainly to large, sparse matrices: the algorithms in~\cite{elgohary2019compressed,elgohary19matrix} are designed for matrices coming from Machine Learning domains; while the ones in~\cite{francisco2018exploiting} are specialised to binary matrices representing the adjacency matrix of web and social graphs.

In this paper we continue the line of research introduced in~\cite{elgohary2019compressed,elgohary19matrix}, called {\em Compressed Linear Algebra} (CLA). These authors use relatively simple compression techniques (e.g., Offset-List Encoding, Run-Length Encoding, Direct Dictionary Coding) preceded by a {\em compression planning} phase that partitions the columns of the input matrix into groups that can be effectively compressed together. Since ML matrices often exhibit hidden correlations (see, for instance, \cite{elgohary19matrix}), the combination of a careful compression planning, which is done only once, together with simple compression techniques yield good compression and fast linear algebra operations. To improve performance, the CLA system also deploys row and column partitioning techniques, so that the compression is cache-friendly and suitable for multithreading. 

We design and experiment new lossless compression schemes for large matrices, which get the best performance when the input matrices are either sparse or contain a relatively small number of {\em distinct} values. A fundamental feature of our contribution is that our lossless compression algorithms guarantee that: 

\begin{itemize}

    \item the compression ratio is bounded in terms of the $k$-th order empirical entropy of the compressed sparse row/value (CSRV) representation of the input matrix; and 
    
    \item the cost of the right and left matrix-vector multiplication is proportional to the size of the {\it compressed} matrix. 
    
\end{itemize}

As we just mentioned, achieving simultaneously a saving in time and space is not new~\cite{elgohary2019compressed,elgohary19matrix,francisco2018exploiting,KourtisGK08}, but to our knowledge our algorithmic approach is the first one achieving bounds for the time and space complexities that match the theoretical limit expressed by the $k$-th order statistical entropy. Given its theoretical properties, our grammar-based algorithm could be also used not only as stand-alone compression tools for matrices but also as a new powerful compression option within the CLA framework, or a similar system, in place of its simpler compressors. 

\smallbreak Technically speaking, the CSRV representation of a matrix is a simple modification to the well-known compressed sparse row (CSR) representation~\cite{Saad:2003}, which is more effective than CSR when the input matrix contains relatively few distinct values. In Section~\ref{sec:proposal} we show that we can compress this representation using a grammar compressor (see e.g.~\cite{tit/KiefferY00}) so that we can later compute the right and left matrix-vector multiplication by working directly upon the compressed matrix, and within time and working space proportional to the compressed size of that matrix. We tested our proposal in practice with a prototype described in Section~\ref{sec:implementation} using the RePair~\cite{tois/MoffatP20} grammar compressor over seven matrices referring to real ML problems. In terms of attained compression ratios, the experiments show that our tool is clearly superior to \gzip , and that it is usually within 20\% of \xz; but, in addition, our solution offers support for linear algebra operations directly over the compressed file whereas \gzip and \xz cannot. 

To measure the efficiency of the matrix multiplication algorithms, we tested a sequence of left and right vector-matrix multiplications and found that the peak memory usage for our algorithms is within 4.11\% for a single thread and within 6.14\% for 16 threads of the size of the uncompressed matrix. These results confirm the theoretical finding that grammar compression can indeed save a significant amount of space during the computation, and therefore allows us to work with larger data sets in internal memory.

In the second part of the paper, encouraged by these results we add an algorithmic step to our grammar-based compression scheme that obtains an even greater space saving. As pointed out in~\cite{elgohary2019compressed}, ML matrices often exhibit correlations between columns; this phenomenon is likely to make the same combination of values appear in the same columns in multiple rows. Most compressors are able to exploit the presence of identical values only when they occur in contiguous columns. Nonetheless, in real-world data sets correlated columns often appear far apart from each other. For this reason, the matrix compression scheme of CLA~\cite{elgohary2019compressed} features a preliminary step aimed at discovering groups of correlated columns; at a later stage, such groups get compressed independently one another, possibly choosing a different compressor for each group. We hence study the problem of column reordering under the hypothesis that the subsequent compression phase is implemented via a grammar compressor. The problem of column reordering in binary, categorical and general matrices has attracted a lot of interest in the past because of its applications to compressing tables arising from several contexts, such as data warehouses \cite{BFG03,ChakrabartiPMF04,VO2007273}, biological experiments~\cite{apostolico08}, mobile data \cite{HanL16}, and graph DBs \cite{JohnsonKCKV04}, just to cite a few. Discovering dependency relations among matrix columns and finding  the column order that guarantees the smallest compressed output is an NP-hard problem in its general form (cf.~e.g.~\cite{BFG03}). For this reason, all the papers above use heuristics to efficiently find appropriate column permutations. In all those cases, the first key step lies in the definition of a proper measure of {\em column similarity} accounting for the special features of the problem and of the compressor at hand. 

In Section~\ref{sec:colreoder}, we present a column-simila\-ri\-ty score designed to work with our previous lossless grammar-based compressors for matrices, which can be computed efficiently. Then, we describe four new algorithms for column reordering that hinge on this score and, in order to boost their performance, we apply them to blocks of rows which are finally compressed individually. We test the effectiveness of this combination over the same seven ML matrices mentioned before. Experimental figures show that, with a modest preprocessing time and no worsening of the running time, we can get a further reduction of up to 16\% of the peak memory usage during matrix-vector multiplication.

As a final contribution of this paper, we compare our matrix compressor against the Compressed Linear Algebra (CLA) approach, which constitute the state-of-the-art in this setting~\cite{elgohary2019compressed,elgohary19matrix}. Experiments show that, in terms of compression, CLA is less effective than our approach over 6 matrices (out of the 7 tested), with a (absolute) space improvement of up to 10\%; in terms of running time, CLA is always at least twice slower than one version of our compressor and at least three times slower than another version of our compressor. Note that these results were obtained using 16 threads for our compressor, whereas CLA is designed to use all the available threads (the testing machine supports up to 80 independent threads). 

Summing up, our experiments show that: (1) our grammar-based compressors for matrices are indeed able to achieve a better space reduction than CLA, which is not surprising given that they are provably able to compress their input up to the $k$-th order statistical entropy, and (2) the theoretical results ensuring that the number of operations is bounded by the size of the compressed matrix translate to algorithms that are also faster in practice to compute matrix operations. As a final note, we point out that most of the fine-tuning techniques adopted for CLA in~\cite{elgohary2019compressed,elgohary19matrix} are not specific to a particular compressor, hence our grammar-based compressors could undergo the same engineering steps; this also suggests that our compressors could be adopted not only as stand-alone compression tools for matrices but also as a new powerful compression option within the CLA framework.

\subsection{Transparency and Reproducibility}

All source files of our algorithms, as well as the scripts to reproduce the experimental results, are available at the repository \url{https://gitlab.com/manzai/mm-repair}. The data sets are available at the public Kaggle repository~\cite{KaggleMLM}.

\section{The Compressed Sparse Row/Value Representation}\label{sec:vcr}

Given a matrix $M \in\R^{n,m}$ with $n$ rows and $m$ columns, the compressed sparse row (CSR) representation~\cite{Saad:2003} is a classical scheme taking advantage of the matrix sparsity. If the matrix $M$ contains $t$ non-zero elements, the CSR representation consists of 1) a length-$t$ array {\sf nz} listing the non-zero elements row-by-row; 2) a length-$t$ array {\sf idx} storing for each element in {\sf nz} its column index; 3) a length-$n$ array {\sf first} such that $\mathsf{first}[1]=0$, and $\mathsf{first}[i]$ with $2 \leq i \leq n$ equals the number of non-zero terms in the first $i-1$ rows (this information is of interest for partitioning the elements of {\sf nz} by rows). 

If the number of {\em distinct} non-zero values is relatively small, then it is more space efficient to introduce an array $V[1,k]$ containing only the distinct non-zero elements of $M$ and to store in {\sf nz} not the actual non-zero values but their indices in $V$. If there are, say, fewer than $2^{16}$ distinct non-zero elements, then each entry in {\sf nz} only takes 2 bytes instead of the 8 bytes of a {\sf double}: this saving can more than compensate for the extra cost of storing the array~$V$. This representation as a whole is called CSR-IV in~\cite{KourtisGK08}.

In this paper we introduce a new representation, called {\em Compressed Sparse Row/Value} (CSRV), by making two minor modifications to the above scheme. Firstly, we combine the two length-$t$ arrays {\sf nz} and {\sf idx} in a single vector of pairs $S$, such that for $i=1,\ldots,t$, entry $S[i]$ contains the pair of integers $(\mathsf{nz}[i],\mathsf{idx}[i])$. Secondly, instead of storing a separate array $\mathsf{first}$ we include its information in $S$ by storing a special symbol $\sz{i}$ immediately after the last non-zero entry of each row. As a result, the array $S$ now has length $t+n$ and can be seen as obtained by scanning the matrix $M$ row-by-row: for each entry $M[i][j]\neq 0$ we append to $S$ the pair $\sy{\ell}{j}$, where $\ell$ is the index in $V$ such that $V[\ell] = M[i][j]$; in addition, we append to $S$ the special symbol $\sz{i}$ every time the parsing reaches the end of a row. Figure~\ref{fig:vc} reports an example in which the elements of $V$ are sorted according to their size, but actually any other ordering (or no ordering at all) would have worked equally well. Also, the elements of $S$ within the same row can be reordered without loss of information; this latter property will be used in Section~\ref{sec:colreoder} to improve compression. 

\ignore{
Given a matrix $M \in\R^{n,m}$ with $n$ rows and $m$ columns, its \emph{value/column representation} consists of two simple data structures:

\begin{itemize}
    \item an array $V[1,k]$ containing the distinct non-zero elements appearing in the matrix $M$;
    \item a string $S$ built by scanning the matrix $M$ row-by-row rightwards. For each entry $M[i][j]=a_{ij} \neq 0$ at row $i$, we append to $S$ the symbol $\sy{\ell}{j}$, where $\ell$ is the index in $V$ such that $V[\ell] = a_{ij}$, and $j$ is the column at row $i$ where that symbol occurs. In addition, to distinguish the (variable length) encoding of $M$'s rows, we append to $S$ a special (delimiter) symbol $\sz{i}$ every time the parsing reaches the end of a row.   
\end{itemize}}

\begin{figure*}[t!]
	\begin{minipage}{.45\linewidth}
		\centering
		$$\left[\arraycolsep=5pt \begin{array}{rrrrr}
			1.2 & 3.4 & 5.6 & 0   & 2.3\\
			2.3 & 0   & 2.3 & 4.5 & 1.7\\
			1.2 & 3.4 & 2.3 & 4.5 & 0 \\
			3.4 & 0   & 5.6 & 0   & 2.3\\
			2.3 & 0   & 2.3 & 4.5 & 0  \\
			1.2 & 3.4 & 2.3 & 4.5 & 3.4\\
		\end{array} \right]$$
	\end{minipage}%
	\begin{minipage}{.55\linewidth}
		\begin{align*}
			V \;=\;  &[1.2\quad 1.7\quad 2.3\quad 3.4\quad 4.5\quad 5.6]\\[5pt]
			S \;=\;  &\sy{1}{1}\; \sy{4}{2}\; \sy{6}{3}\; \sy{3}{5}\; \sz{1}\;
			\sy{3}{1}\; \sy{3}{3}\; \sy{5}{4}\; \sy{2}{5}\; \sz{2}\\
			&\sy{1}{1}\; \sy{4}{2}\; \sy{3}{3}\; \sy{5}{4}\; \sz{3}\;
			\sy{4}{1}\; \sy{6}{3}\; \sy{3}{5}\; \sz{4}\\
			& \sy{3}{1}\; \sy{3}{3}\; \sy{5}{4}\; \sz{5}\;
			\sy{1}{1}\; \sy{4}{2}\; \sy{3}{3}\; \sy{5}{4}\; \sy{4}{5}\;\sz{6}
		\end{align*}
	\end{minipage}
	\caption{A matrix and its CSRV representation. In the array $S$ the symbol $\sy{3}{1}$ stands for an occurrence of the value $V[3]=2.3$ in column 1. Note that the same value in column 3, is represented instead by $\sy{3}{3}$.
		Only the same values in the \emph{same column} are represented by the same pair $\sy{i}{j}$.}
	\label{fig:vc}
\end{figure*}

\ignore{
\noindent See Figure~\ref{fig:vc} for an example, where the elements of $V$ are sorted according to their size, but any ordering (or no ordering at all) would work equally well. In the same vein, we could use other encodings for the delimiter $\sz{i}$, here we adopt the simplest one.

If the matrix $M$ contains $t$ non-zero elements $S$ has length $t+n$ and its alphabet is $\A = \{\sz{i}\} \cup \{ \sy{\ell}{j} \vert 1\leq \ell \leq k,\, 1\leq j \leq m\} \cup \{\sz{r} \}$.

From $S$ and $V$ we can recover the original matrix $M$. We note that our value/column representation is similar to the well-known Compressed Sparse Row (CSR) representation which consists of two arrays: one of length $t$ containing the non-zero values and their column index, and one of length $n$ containing pointers to the starting position of each row in the above array. }

Given the CSRV representation of matrix $M$ and a vector $x[1,m]$, it is straightforward to compute the matrix-vector multiplication $y=Mx$ with a single scan of $S$. To begin with, we initialise the vector $y[1,n]$ to zero. Then, during the scan of row $i$, when we encounter the pair $\sy{\ell}{j}$ we add the value $V[\ell] \cdot x[j]$ to the entry $y[i]$. The occurrences of the symbol $\sz{i}$ allow us to keep track of the current row. 
 We can similarly compute with a single scan of $S$ the left-multiplication $x^t = y^t M$: firstly, we initialise $x[1,m]$ to zero; then, during the scan of row $i$, when we encounter the pair $\sy{\ell}{j}$ we add the value $y[i] \cdot V[\ell]$ to the entry $x[j]$. Hence, either right and left multiplications can be computed in $\Oh(|S|) = \Oh(n+t)$ time.  Hereinafter we use the notation $\CSRV$ to denote the CSRV representation outlined above.

\section{Grammar-Compressed Matrices}
\label{sec:proposal}

In this section we show how to compress the CSRV representation \CSRV of a matrix $M$. Whenever such representation offers room for further compression, we provably get a reduction in both the space occupancy and in the cost of performing the left and the right vector-to-matrix multiplication operations. 

Recall that a grammar-compressed representation for a string $T$ over an alphabet of terminal symbols $\A$ is a context-free grammar that generates only~$T$~\cite{tit/CharikarLLPPSS05}. For the sake of simplicity, we assume that the grammar is a so-called {\em straight-line program}~\cite{gcc/Lohrey12} (SLP), that is, it consists of a set of rules of the form $L_i \to R_{i_1}R_{i_2}$, where $L_i$ is a nonterminal and each of $R_{i_1}$ and $R_{i_2}$ can be either a terminal (i.e., an element of the base alphabet $\A$), or a nonterminal.  The grammar generates only $T$, implying that each nonterminal appears as the left-hand side of a single rule, so that one can identify each rule with the nonterminal on its left-hand size.
Given a nonterminal $N_j$, its {\em expansion}, denoted by $\Exp(N_j)$, is defined as the (unique) sequence we obtain by repeatedly applying the substitution rules of the SLP grammar until we are left with a string over~$\A$. Thus one can leverage a SLP to represent $T$ as a succinct sequence $C$ of nonterminals; when needed, $T$ can be obtained from $C$ by evaluating the expansion of its nonterminals.

%With a little abuse of notation we define the expansion also for terminal symbols defining $\Exp(\sigma)=\sigma$ for all $\sigma\in\A$. 

The output of the grammar compressor is a set of rules and a special nonterminal whose expansion generates only the input string~$T$. If the grammar has $q$ rules, and therefore $q$ nonterminals $N_1, \ldots, N_q$, we can number them so that if $N_i$ appears in the right-hand side of $N_j$, then $i<j$. 

One can define the size of a grammar as the sum of the lengths of the right-hand sides of the rules. The same text $T$ can be generated by many different grammars, and finding the smallest one is NP-complete \cite{tit/CharikarLLPPSS05,jacm/StorerS82}. Yet, the compressors producing irreducible grammars, among them
{\sf Greedy}, {\sf LongestMatch}~\cite{tit/KiefferY00}, {\sf RePair}~\cite{larsson2000off}, and {\sf Sequential}~\cite{tit/KiefferY00}, are guaranteed to produce an output whose size in bounded by $|T| H_k(T) + o(|T| H_k(T))$ bits for any $k\in o(\log_\sigma |T|)$, where $\sigma$ is the size of the input alphabet and $H_k(T)$ is the order-$k$ statistical entropy of the input $T$~\cite{tit/OchoaN19}. Up to lower order terms, then, these grammar compression algorithms are as good as the best statistical encoders that compress the input on the basis of the frequencies of $k$-tuples of symbols. Grammar compressors are very effective also for compressing strings with many repetitions: in this case their output size can be within a logarithmic factor from the output of the best compressors based on LZ-parsing; see~\cite{csur/Navarro21a} for details.

% Grammar compressors got interest in the literature because they allow for direct access to the compressed data in logarithmic time. We exploit this property in compressed-matrix multiplication. ** SEMMAI SI COPIA NELL'INTRO.

To compress a CSRV representation \CSRV we apply a grammar compressor to the sequence $S$. We modify the compressor so that it never uses the special terminal symbol $\sz{i}$ in any rule. This guarantees that the expansion of any nonterminal $N_k$ only contains pairs $\sy{i}{j}$. As a result, the output of the grammar compressor applied to $S$ consists of a set of rules $\Ru$ and a string
\begin{equation}
    \Cset = N_{i_1}\sz{1} N_{i_2} \sz{2} \cdots N_{i_n} \sz{n}
\end{equation}
such that each $N_{i_j}$ is a nonterminal whose expansion is the sequence of pairs representing the non-zero elements of row $j$. In the same sense, the expansion of the string $\Cset$ (i.e., expanding each of its nonterminals) is the sequence $S$. An example of a grammar representing the string $S$ of Figure~\ref{fig:vc} is given in Figure~\ref{fig:grammar}. In the following we write $(\Cset,\Ru,V)$ to denote the grammar representation of (the CSRV representation of) a matrix $M$. 

% this is an implemetation detail
% In our prototype we build the grammar using RePair~\cite{larsson2000off}, a linear time and linear space algorithm which guarantees to build an SLP grammar with provable compression performance~\cite{tit/OchoaN19}. However, our results are independent of the tool used for the construction of the grammar representing~$S$.

\begin{figure}[t!]
	\begin{align*}
		\Ru \;=\;\{ & N_1 \to \sy{3}{3}\, \sy{5}{4} \quad
		N_2 \to \sy{1}{1}\, \sy{4}{2}\quad
		N_3 \to \sy{3}{1}\, N_1\\
		& N_4  \to \sy{6}{3}\, \sy{3}{5}\quad
		N_5  \to       N_2\, N_4 \qquad
		N_6  \to       N_3\, \sy{2}{5}\\
		& N_7  \to       N_2\, N_1 \qquad 
		N_8  \to \sy{4}{1}\, N_4\quad\;\;
		N_9  \to       N_7\, \sy{4}{5}\;\} \\ \\
		{\Cset} \;=\; & N_5\, \sz{1}\, N_6\, \sz{2}\, N_7\, \sz{3}\, 
		N_8\, \sz{4}\, N_3\, \sz{5}\, N_9\, \sz{6}
	\end{align*}
	\caption{The set of rules $\Ru$ and the final string $\Cset$ whose expansion is the sequence $S$ from Figure~\ref{fig:vc}.}
	\label{fig:grammar}
\end{figure}

\subsection{Right Multiplication for Grammar-Compressed Matrices}\label{sec:rmultiplication}

In this section we show that, given a grammar representation \GRAMM of a matrix $M$, we can compute the right multiplication $y=Mx$ in $\Oh(|\Ru|)$ time using $\Oh(|\Ru|)$ words of auxiliary space. In the following we use $S$ to denote the expansion of $\Cset$, so that \CSRV is the CSRV representation of $M$.

\begin{definition}
	Given a vector $x[1,m]$ and a pair $\sy{\ell}{j}\in S$ we define 
	$$
		\Val_x(\sy{\ell}{j}) = V[\ell] \cdot x[j]; 
	$$
	(recall that the pair $\sy{\ell}{j}$ represents the value $V[\ell]$ stored in column~$j$ of matrix $M$). 
	Similarly, for a nonterminal $N_i$ whose expansion is 
	$\sy{\ell_1}{j_1}$ $\sy{\ell_2}{j_2} \cdots \sy{\ell_h}{j_h}$ we define
	\begin{equation}
		\Val_x(N_i) = \sum_{k=1}^h \Val_x(\sy{\ell_k}{j_k}) = \sum_{k=1}^h V[\ell_k] \; x[j_k].
	\end{equation}
\end{definition}

\noindent From the above definition we immediately get

\begin{lemma}\label{lemma:val}
	If the grammar contains the rule $N_i \to AB$, then $\Val_x(N_i) = \Val_x(A) + \Val_x(B)$.\qed
\end{lemma}

\begin{lemma}\label{lemma:y=mx}
	Given the representation \GRAMM of a  matrix $M\in\RsupMN$ with $\Cset = N_{i_1}\sz{1} \cdots N_{i_n} \sz{n}$, if $y=Mx$ then it holds that $y[r] =  \Val_x(N_{i_r})$, for $r=1,\ldots, n$.
\end{lemma}

\begin{proof}
    We have $y[r] = \sum_{i=1}^m  M[r][i]\cdot  x[i]$. By construction, the expansion of the nonterminal $N_{i_r}$ is the sequence of pairs $\sy{\ell_1}{j_1} \cdots \sy{\ell_h}{j_h}$ representing all the non-zero elements of row~$r$ where, for $k=1,\ldots,h$, $\ell_k$ denotes the position in $V$ containing the value $M[r][j_k]$. Thus 
\begin{align*}
 y[r] & = \sum_{k=1}^h M[r][j_k] \cdot x[j_k]\\
      & = \sum_{k=1}^h V[\ell_k] \cdot x[j_k]
      \; = \; \Val_x(N_{i_r}).
\end{align*}
\end{proof}

% Suppose we want to compute the right multiplication $y=Mx$. If the grammar compressor applied to $M$ produces the final string $C= N_{i_1}\sz{1} N_{i_2} \sz{2} \cdots N_{i_n} \sz{n}$  then for $\ell=1,\ldots,n$ it is $y[\ell] = \Val_x(N_{i_\ell})$.

\begin{theorem}\label{theo:y=mx}
	Given the representation \GRAMM of a matrix $M\in\R^{n\times m}$ and a vector $x \in \R^m$, we can compute $y=Mx$ in $\Oh(|\Cset|+|\Ru|)$ time using $\Oh(|\Ru|)$ words of auxiliary space.
\end{theorem}

\begin{proof}
To compute $y=Mx$, we introduce an auxiliary array $W[1,q]$, where $q=|\Ru|$, such that $W[i]=\Val_x(N_i)$. Because of Lemma~\ref{lemma:val} and of the rule ordering, we can fill $W$ with a single pass over $\Ru$ in time $\Oh(q)$: the value $W[i]=\Val_x(N_i)$ is the sum of two terms that can be either of the form $\Val_x(\sy{h}{k})$ or $\Val_x(N_j)$ with $j<i$. In the former case  $\Val_x(\sy{h}{k}) = V[h]\cdot x[k]$; in the latter case $\Val_x(N_j) = W[j]$ for some already computed entry, since $j<i$. One may indeed observe that $N_i$'s are ranked by the time when they get computed. After filling $W$, we use Lemma~\ref{lemma:y=mx} to get the components of the output vector~$y$.
\end{proof}

\subsection{Left multiplication for grammar-compressed matrices}
\label{sec:lmultiplication}

We now show that, given the grammar representation \GRAMM of a matrix $M$, we can compute the left multiplication $x^t = y^t M$ with an algorithm symmetrical to the one for the right multiplication and within the same time and space bounds. %In the following we use $S$ to denote the expansion of $\Cset$, so that \CSRV denotes the CSRV representation of $M$. 

\begin{definition}
	For any $\sy{\ell}{j}\in S$ we define $\Reach(\sy{\ell}{j})$ as the set of rows whose CSRV representation contains~$\sy{\ell}{j}$. Note that $k\in\Reach(\sy{\ell}{j})$ if, and only if, the expansion of the nonterminal $N_{i_k}\in\Cset$ contains the pair  $\sy{\ell}{j}$ or, equivalently, $M[k][j] = V[\ell]$. 
\end{definition}

% Note that  $\ell\in\Reach(\sy{i}{j})$ if and only if $M[\ell][j] = V[i]$.

For the example in Figure~\ref{fig:vc}, we have $\Reach(\sy{1}{1}) = \{1,3,6\}$ since $\sy{1}{1}$ represents the value 1.2  that appears in column 1 of those three rows. Similarly, $\Reach(\sy{3}{1}) = \{2,5\}$.  

\begin{definition}
	Given a vector $y[1,n]$, for any $\sy{i}{j}\in S$ we define $\Sum_y(\sy{i}{j})$ as 
	$$
	\Sum_y(\sy{i}{j}) = \sum\nolimits_{k\in\Reach(\sy{i}{j})} y[k]
	$$
\end{definition}

\begin{lemma}\label{lemma:xj}
	Given the CSRV representation $\CSRV$ of matrix $M\in\R^{n\times m}$, let $S'$ be the set of distinct symbols in $S$ (i.e., without duplicates). If $x^t = y^t M$ then, for $j=1,\ldots,m$, it holds
	$$
	x[j] = \sum\nolimits_{\sy{i}{j}\in S'} V[i] \cdot \Sum_y(\sy{i}{j})
	$$
	(note the summation involves only pairs in $S'$ whose second component is~$j$).  
\end{lemma}

\begin{proof}
	Since
	$$x[j] = \sum\nolimits_{\ell=1}^n  y[\ell] \cdot M[\ell][j],$$
	the value $x[j]$ depends only upon the non-zero elements in column~$j$. Each nonzero in column $j$ gets represented by a symbol $\sy{i}{j}$ and has its corresponding value encoded by some entry $V[i]$. If $\sy{i}{j}$ occurs at row $r$ in column $j$, then $y[r]$ is multiplied by $V[i]$, and this holds for all rows containing $\sy{i}{j}$. One can aggregate these multiplications and write them as $V[i] \cdot \Sum_y(\sy{i}{j})$. 
	The lemma follows by iterating this argument over all distinct non-null values $V[i]$ occurring in column $j$, and therefore over all pairs $\sy{i}{j}\in S'$.
\end{proof}

We now show that the notions of $\Reach$ and $\Sum$ can be naturally extended to nonterminals.

\begin{definition}
   Given the representation \GRAMM of a matrix $M\in\R^{n\times m}$,
	for each nonterminal $N_j$ we define $\Reach(N_j)$ as the set of row indices $\ell$ such that $N_j$ appears in the expansion of $N_{i_\ell}$. In other words, $\Reach(N_j)$ denotes the rows whose compression makes use of $N_j$. We also define $\Sum_y(N_j) = \sum\nolimits_{\ell\in\Reach(N_j)} y[\ell]$. 
\end{definition}

In the following we make the natural assumption that the grammar does not contain {\em useless} rules, that is, if the grammar contains the rule $N_i \to AB$, then $N_i$ appears in the right-hand side of some other rule (whose left-hand side will be some $N_j$ with $j>i$), or $N_i$ appears in the final string $\Cset$ (or both). 

\begin{lemma} \label{lemma:sum}
	For any symbol $\alpha$ (terminal or nonterminal), let $\Ru_\alpha$ denote the set of nonterminals $N_j$'s such that their defining rule $N_j\to AB$ contains $\alpha$ in their right-hand side (i.e., $A = \alpha $ or $B = \alpha$), and let $\Xx_\alpha$ denote the set of row indices $\ell$ such that $N_{i_\ell}=\alpha$ (hence $\ell\in \Xx_\alpha$ iff the expansion of $\alpha$ coincides with the $\ell$-th row). Then,
	\begin{equation}\label{eq:+delta}
	\Sum_y(\alpha) = \sum\nolimits_{N_j \in R_\alpha}\!\Sum_y(N_j)
	\;\;+\;\; \sum\nolimits_{\ell\in \Xx_\alpha}  y[\ell].
	\end{equation}
\end{lemma}

\begin{proof}
   Since each occurrence of $\alpha$ is either the right-hand side of a single rule, or coincides with some $N_{i_\alpha}$, we have
   \begin{equation*}
       \Reach(\alpha) = \left\{\bigcup\nolimits_{N_j\in\Ru_\alpha}\!\Reach(N_j)\right\} \;\;\bigcup_\alpha\;\;
       \Xx_\alpha
   \end{equation*}
   and the lemma follows by induction on the number of steps in the derivation of $\alpha$. 
\end{proof}

In view of Lemma~\ref{lemma:xj}, to compute $x^t = y^t M$, we need to compute $V[i]\cdot\Sum_y(\sy{i}{j})$ for all $\sy{i}{j}\in S'$. To this end we first compute $\Sum_y$ for nonterminals and then we use Lemma~\ref{lemma:sum} to derive the values $\Sum_y(\sy{i}{j})$.

In our implementation we introduce an auxiliary array $W[1,q]$, where $q=|\Ru|$, such that at the end of the computation $W[i]$ contains $\Sum_y(N_i)$. 
More in detail, we initially set $x[1,m]$ to zero, and we set $W[1,q]$ to zero as well, except for the entries $W[i_\ell]$ that we initialise to $y[\ell]$ for every nonterminal $N_{i_\ell}$ in the final string $\Cset$ (this accounts for the terms in the second summation of~\eqref{eq:+delta}). Next, we scan the set of rules {\em backwards} from $q$ to $1$; for every rule $N_j \to AB$ we proceed as follows:
\begin{itemize}
	\item if $A$ (or $B$) is equal to another nonterminal $N_i$ (necessarily with $i<j$) we increase $W[i]$ by the value $W[j]$;
	\item if $A$ (or $B$) is equal to a terminal $\sy{h}{k}$ we increase $x[k]$ by $V[h]\cdot W[j]$. 
\end{itemize}
The crucial observation is that when we reach the rule $N_j \to AB$ we have already computed in $W[j]$ the correct value $\Sum_y(N_j)$ since we have already accounted for all terms in Lemma~\ref{lemma:sum}, namely the nonterminals in the final string $\Cset$ and all rules containing $N_j$ in their right-hand side (by our assumptions these rules will be numbered higher than~$j$). Using our strategy, the value $\Sum_y(N_j)$ is added to $\Sum_y(A)$ and $\Sum_y(B)$, affecting their corresponding values in $W$ if they are nonterminal, or being accumulated in the proper entry of $x$ if they are terminals.

\begin{theorem}\label{theo:x=ym}
	Given the representation \GRAMM of a matrix $M\in\R^{n\times m}$ and a vector $y \in \R^n$, we can compute $x^t=y^t M$ within $\Oh(|\Cset|+|\Ru|)$ time using $\Oh(|\Ru|)$ words of auxiliary space.\qed
\end{theorem}

We point out that we do not require that in the array $S$, which is compressed to $\Cset$ and $\Ru$, the pairs relative to the same row are ordered according to column index, as we arranged them in Figure~\ref{fig:vc}. For helping the compression, we could instead reorder the pairs in other ways: this would not impact upon the design of our multiplication algorithms. In Section~\ref{sec:colreoder}, we analyse the improvement in compression obtained by reordering the columns of $M$ {\em globally}: namely, reordering the elements in each row according to the same permutation. As for future work, we plan to analyse the general problem in which the elements in each row are reordered independently of all other rows.

\section{Implementation and experiments}
\label{sec:implementation}

In this section we describe a prototype of our matrix multiplication algorithm for grammar-compressed matrices. As we will see, the original idea can be refined to get different representations with different time/space trade-offs so that, in the end, we will eventually get a {\em family} of grammar-compression algorithms.

Given a matrix $M \in\RsupMN$ we first build the CSRV representation \CSRV as described in Section~\ref{sec:vcr}. We implemented this representation by storing the sequence $S$ as an array of 32-bit unsigned integers: the symbol $\sz{}$ is encoded by the integer $0$, while the pair $\sy{i}{j}$ is encoded by the integer $1+im+j$ (recall that $0\leq j <m$ is the column index). The entries of $V$ are represented as 8-byte doubles, so the total space usage amounts to $4|S| + 8|V|$ bytes. In the following we call this representation \csrv and we use it as a baseline for our tests. 

To build the grammar representation \GRAMM we compress the 32-bit integer sequence $S$ using the  RePair algorithm~\cite{larsson2000off}, which runs in $O(|S|)$ time, using $O(|S|)$ words of space, and achieves a compression ratio bounded by the high-order statistical entropy of~$S$ (see Sect.~\ref{sec:proposal}). RePair works by repeatedly finding the most frequent pair of consecutive symbols $AB$, replacing all occurrences $AB$ by a new nonterminal $N_i$, and appending the rule $N_i\to AB$ to the current rule set. We modified RePair so that it never builds a rule involving the symbol $\sz{}$, as required by our construction.
RePair stops when there are no pairs of consecutive symbols appearing more than once.  As a consequence, the final  string $\Cset$ has not necessarily the form $N_{i_1}\sz{1} N_{i_2} \sz{2} \cdots N_{i_n} \sz{n}$ discussed in the previous section; RePair's final string $\Cset$ is usually longer and may even include terminals $\sy{i}{j}$. We could add additional rules to obtain a final string  $\Cset$  with exactly $2n$ symbols as above; however, since this does not help compression or running times we use RePair's final string as $\Cset$, adding the (simple) necessary modifications to the multiplication algorithm.

In addition to the final string $\Cset$, RePair produces a set of rules $\Ru$ where, as already mentioned, each rule is represented by a pair of symbols. In its na\"ive representation, RePair outputs $|\Cset|+2|\Ru|$ 32-bit integers overall.\footnote{Notice that for a rule $N_i \rightarrow AB$, we have to encode only $A$ and $B$ because the nonterminals $N_i$ get increasing ids.}  However, this is quite a wasteful representation: if the largest nonterminal is represented by the integer $N_{\max}$, we can represent $\Cset$ and $\Ru$ using packed arrays with entries of $w=1+\left\lfloor \log_2 N_{\max} \right\rfloor$ bits. What is more, some symbols might be more frequent than others in $\Cset$ or $\Ru$, so we can save additional space by using a variable-length representation {via an entropy coder}. Therefore we have experimented the following variants of RePair compression, {which  induce three corresponding variants of our matrix compression algorithm}:

\begin{description}

\item[\reint:] $\Cset$ and $\Ru$ are represented as 32-bit integer arrays. This is the fastest, but less space-efficient representation.

\item[\reiv:] $\Cset$ and $\Ru$ are represented as packed arrays, with entries of $1+\left\lfloor \log_2 N_{\max} \right\rfloor$ bits as described above. In our implementation we used the class {\tt int\_vector} from the sdsl-lite library~\cite{gbmp2014sea}.

\item[\reansiv:] $\Ru$ is represented via a packed array as above, whereas $\Cset$ is compressed using the {\tt ans-fold} entropy coder from~\cite{tois/MoffatP20}.
\end{description}

All the above variants store the array $V$ uncompressed. 
Clearly, more complex representations are possible, offering even larger compression achievements. However, the reader should notice two important points. Firstly, we want to efficiently support matrix-vector multiplication: looking at the algorithms in Section~\ref{sec:proposal} we see that the left multiplication algorithm scans the rules in $\Ru$ backwards, and only a few compressors provide fast right-to-left access to uncompressed data. In addition, the compression of $\Cset$ and $\Ru$ is secondary: we expect the largest saving from the use of the grammar compressor and of the reordering techniques introduced in Section~\ref{sec:colreoder}.

\subsection{Multi-threaded implementation}\label{sec:parallel}

To take advantage of modern multicore architectures, matrix multiplication algorithms usually split the input matrices into blocks. Indeed, most operations on the individual blocks can be easily carried out in parallel on a multi-thread machine. Since for ML matrices the number of observations (rows) is much larger than the number of features (columns), we implemented a representation in which the input matrix gets partitioned into blocks of rows. Given a parameter $b > 1$, a $r\times c$ matrix $M$ is partitioned into $b$ block of size $\lceil r/b\rceil \times c$ (except for the last block might get fewer rows). With this setting, the right multiplication $y=Mx$ consists of $b$ independent right multiplications each one involving a single block. The left multiplication $x^t = y^t M$ also consists of $b$ independent left multiplications followed by a step in which the $b$ resulting row vectors are summed together. 

Our grammar-based representations can be easily adapted to work with distinct blocks of rows. After computing the CSRV representation  \CSRV, we partition the vector $S$ into $b$ subvectors $S_1, \ldots, S_b$, so that $S_i$ contains the encoding of the non-zero elements of the $i$-th row block. We thereby grammar-compress each subvector $S_i$ using RePair; the resulting string $\Cset_i$ and rule set $\Ru_i$ are then further compressed as described before. Notice that the value array $V$ is unique and shared by all matrix blocks.

\subsection{Some experimental figures}

% The last seven columns report the size of the matrix compressed with different tools described in the text. Sizes are given as percentage {of the ratio between the size of the compressed matrix over the size of the uncompressed full matrix representation} (rows$\times$cols$\times$8 bytes, not reported here); a smaller percentage corresponds to a better compression.

\begin{table*}[t]

\caption{Matrices used in our experiments and the compression ratio achieved by the tools described in the text; a smaller percentage corresponds to a better compression. The column {\sl nonzero} reports the percentage of non-zero elements over the total, while column $\#|\mbox{\sl nonzero}|$ reports the number of {\em distinct} non-zero values.}\label{table:dataset}

%	\centering
	\setlength{\tabcolsep}{6pt} % Default value: 6pt
	\renewcommand{\arraystretch}{1.2} % Default value: 1
	\scalebox{0.8}{\begin{tabular}{l||r|r||r|r||r|r||r||r|r|r}
		\hline
		matrix & \multicolumn{1}{c|}{rows} & cols & \multicolumn{1}{c|}{\sl nonzeros}& \multicolumn{1}{c||}{$\#|\mbox{\sl nonzeros}|$}  & \multicolumn{1}{c|}{\gzip}  & \multicolumn{1}{c||}{\xz} & \multicolumn{1}{c||}{\csrv} & \multicolumn{1}{c|}{\reint} &\multicolumn{1}{c|}{\reiv} &
		\multicolumn{1}{c}{\reansiv} \\ \hline
Susy     & 5,000,000 &18 & 98.82\% & 20,352,142& 53.27\% & 43.94\% & 74.80\% & 74.80\% & 69.91\% & 66.63 \% \\
Higgs    &11,000,000 &28 & 92.11\% &  8,083,943& 48.38\% & 31.47\% & 50.46\% & 46.91\% & 41.38\% & 38.05 \% \\
Airline78&14,462,943 &29 & 72.66\% &      7,794& 13.27\% &  7.01\% & 38.06\% & 14.84\% & 11.13\% &  9.27 \% \\
Covtype  &   581,012 &54 & 22.00\% &      6,682&  6.25\% &  3.34\% & 11.95\% &  7.21\% &  4.52\% &  3.87 \% \\
Census   & 2,458,285 &68 & 43.03\% &         45&  5.54\% &  2.79\% & 22.25\% &  3.24\% &  2.02\% &  1.53 \% \\
Optical  &   325,834 &174& 97.50\% &    897,176& 53.54\% & 27.13\% & 50.62\% & 40.70\% & 35.81\% & 34.31 \% \\
Mnist2m  & 2,000,000 &784& 25.25\% &        255&  6.46\% &  4.25\% & 12.69\% &  7.47\% &  5.84\% &  5.33 \% \\
\hline
\end{tabular}}
\end{table*}

We executed all our experiments on a machine equipped with 80 Intel(R) Xeon Gold 6230 CPUs running @ 2.10 GHz, with 360 GB of RAM. We measured running times and peak space usages with the Unix tool {\tt time}. Table~\ref{table:dataset} reports the features of some matrices from some real {ML} problems obtained from the UCI repository~\cite{UCI} except for  \mnistxm~\cite{MNIST} and \airlinexx~\cite{Airline}. For uniformity’s sake, we represent the entries of all matrices as 8-byte doubles, so the uncompressed and full representation of a matrix takes a total of rows $\times$ cols $\times$ 8 bytes. If such representation is compressed with \gzip and \xz, with their default compression level, the resulting compressed files have the sizes reported in columns 6 and 7 of Table~\ref{table:dataset}. Column 8 reports the size of the \csrv representation, whilst the last three columns report the sizes of the three variants of our RePair compressor described above. All sizes are given as percentage {of the ratio between the size of the compressed matrix over the size of the uncompressed matrix representation} (rows\, $\times$\, cols\, $\times$\, 8 bytes), hence a smaller percentage corresponds to a better compression. 

We emphasise that some of the matrices, namely \susy, \higgs, and \optical, are not really sparse, having more than 92\% non-zero elements. The classical CSR representation, where each non-zero entry takes 12 bytes, would take more space than the uncompressed and full representation. The \csrv representation, that takes advantage of repeated values, is already obtaining some compression; in particular for \optical, which has fewer distinct nonzeros, \csrv  has got a reduced space footprint compared to \gzip. Further space reduction is obtained by our grammar-based compressors. This shows that our techniques are not just suitable for sparse matrices, but they provide some advantages on a larger class of matrices with some structure. 

The comparison between the \csrv and \reint output sizes {is of interest to get} some indication of the effectiveness of grammar compression. At one extreme, we see that \reint does not provide for \susy any additional compression to the \csrv representation, suggesting that there are not many pairs of adjacent non-zero values occurring many times in different rows. At the other extreme, we see that \reint provides for \census a six-fold better compression, and \reiv and \reansiv achieve a compression even better than the state of the art tool~\xz. Moreover, we see that our most sophisticated encoder, \reansiv, is significantly better than \gzip, with the only exception of \susy.

Let us now turn our attention to our main interest, namely reducing {\em both} space usage and running time for the matrix multiplication operations. Standard compressors, like \gzip and \xz, need to {fully decompress the compressed matrix representation in order to do any operation on it.} Hence, the cost of any operation is at least proportional to the size of the {\em uncompressed} matrix; conversely, in the previous section we proved that using grammar compression left and right multiplications can be carried out in time proportional to the size of the {\em compressed} matrix. To measure {the practical impact of this theoretical result}, we considered 500 iterations of the computation
\begin{equation}\label{eq:CG}
  y_i = M\, x_i,\quad z^t_i = y_i^t\, M,\quad x_{i+1} = \frac{z_i}{ \Vert z_i \Vert_\infty}  
\end{equation}
where $\Vert z_i \Vert_\infty$ is the largest modulus of the components of~$z_i$. The above computation consists of 500 alternated left and right matrix multiplications and mimics, e.g., the most costly operations of conjugate gradient method used for least square computations. 

\begin{figure*}[h]
\begin{subfigure}{0.45\textwidth}
\includegraphics[width=1.00\linewidth]{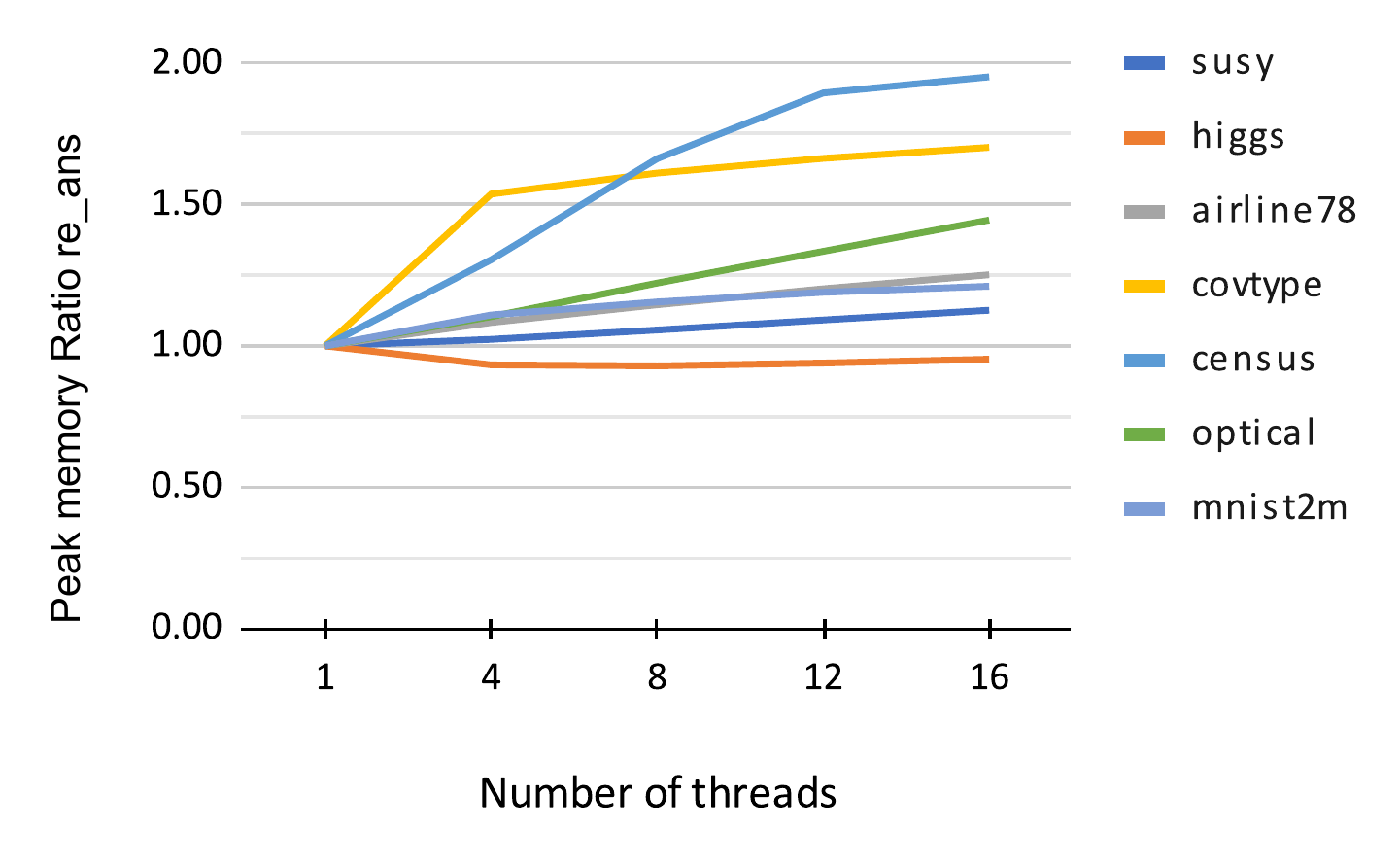}
\end{subfigure}\hspace{1cm}
\begin{subfigure}{0.45\textwidth}
\includegraphics[width=1.00\linewidth]{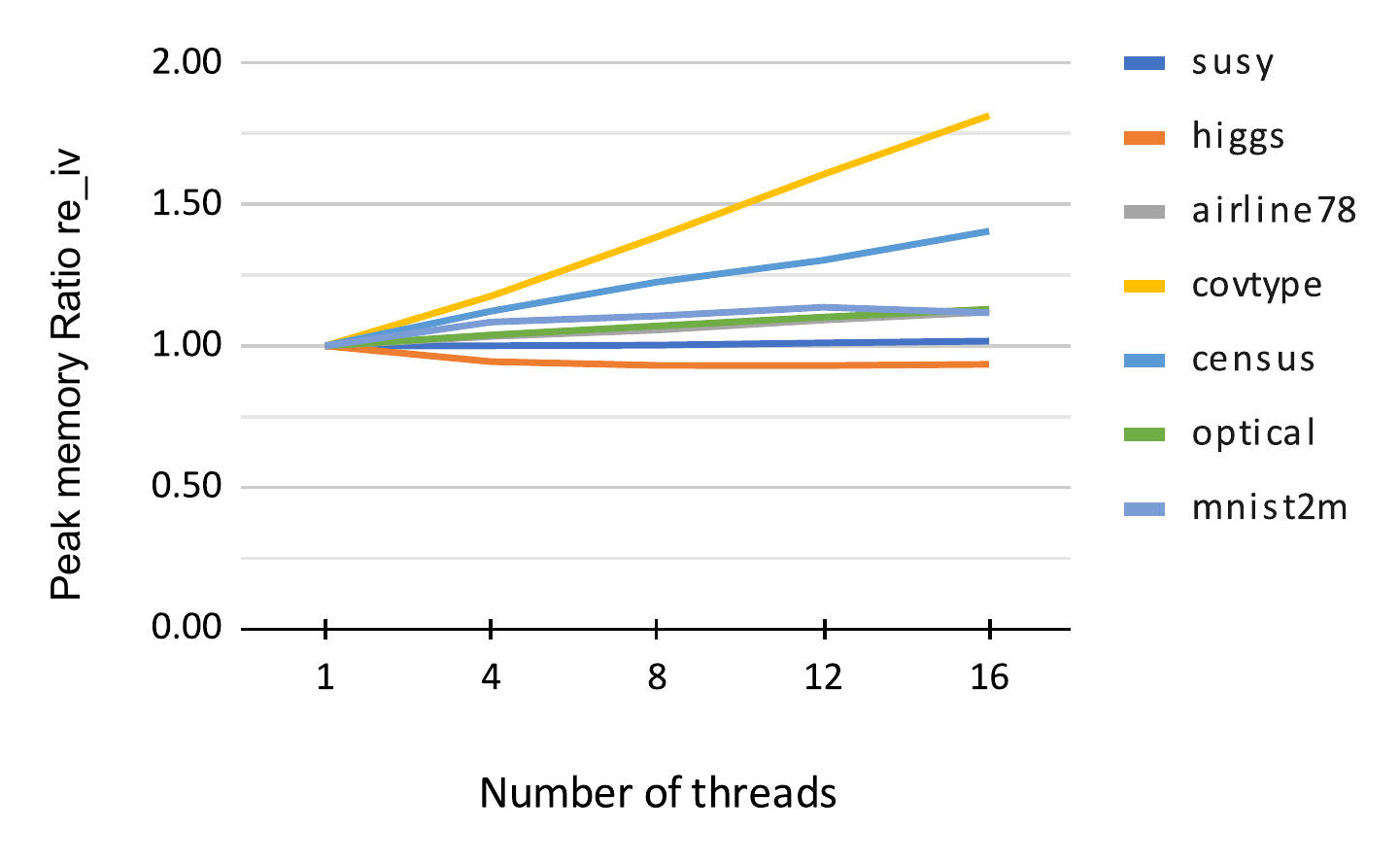}
\end{subfigure}
\begin{subfigure}{0.45\textwidth}
\includegraphics[width=1.00\linewidth]{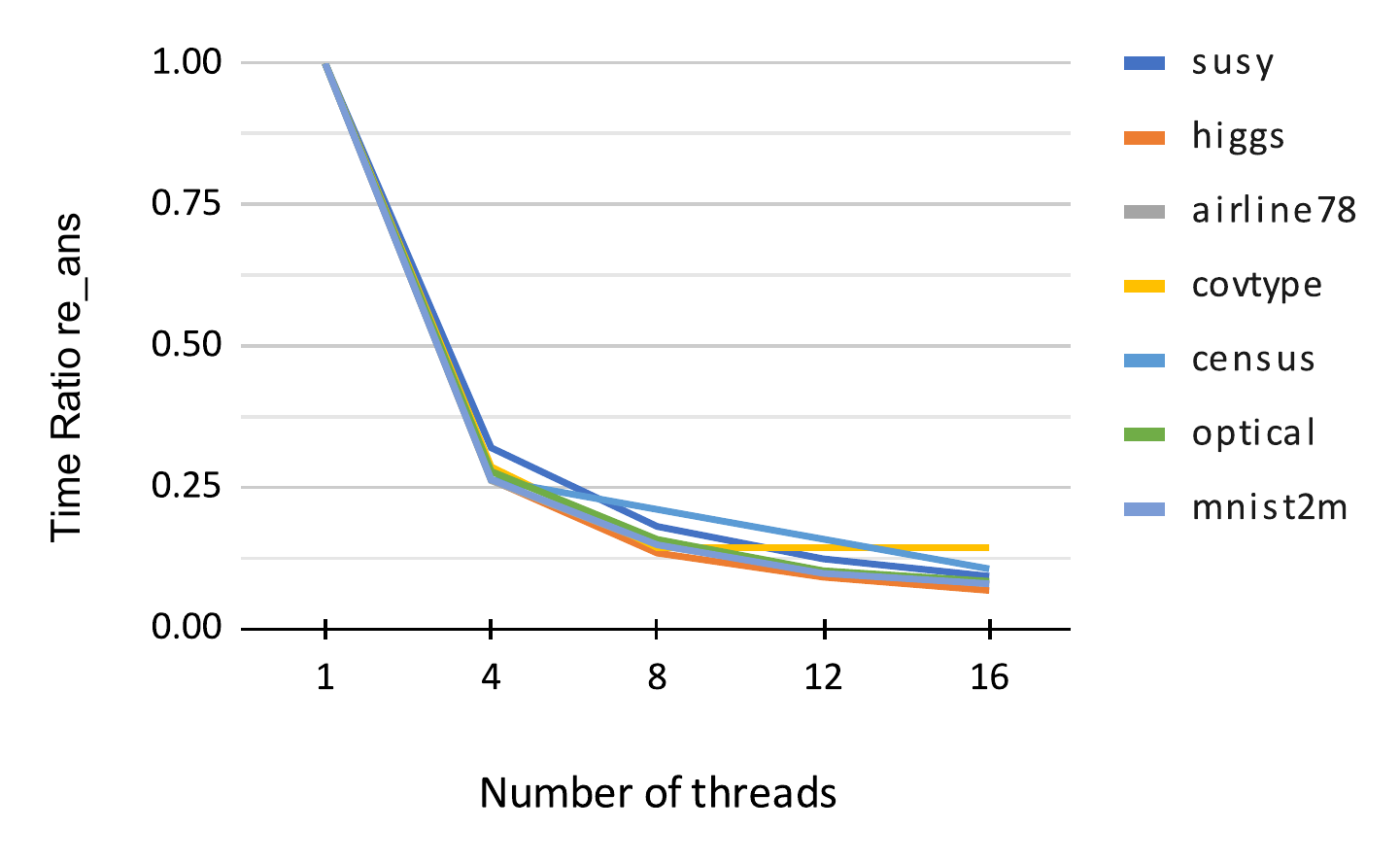}
\end{subfigure}\hspace{1cm}
\begin{subfigure}{0.45\textwidth}
\includegraphics[width=1.00\linewidth]{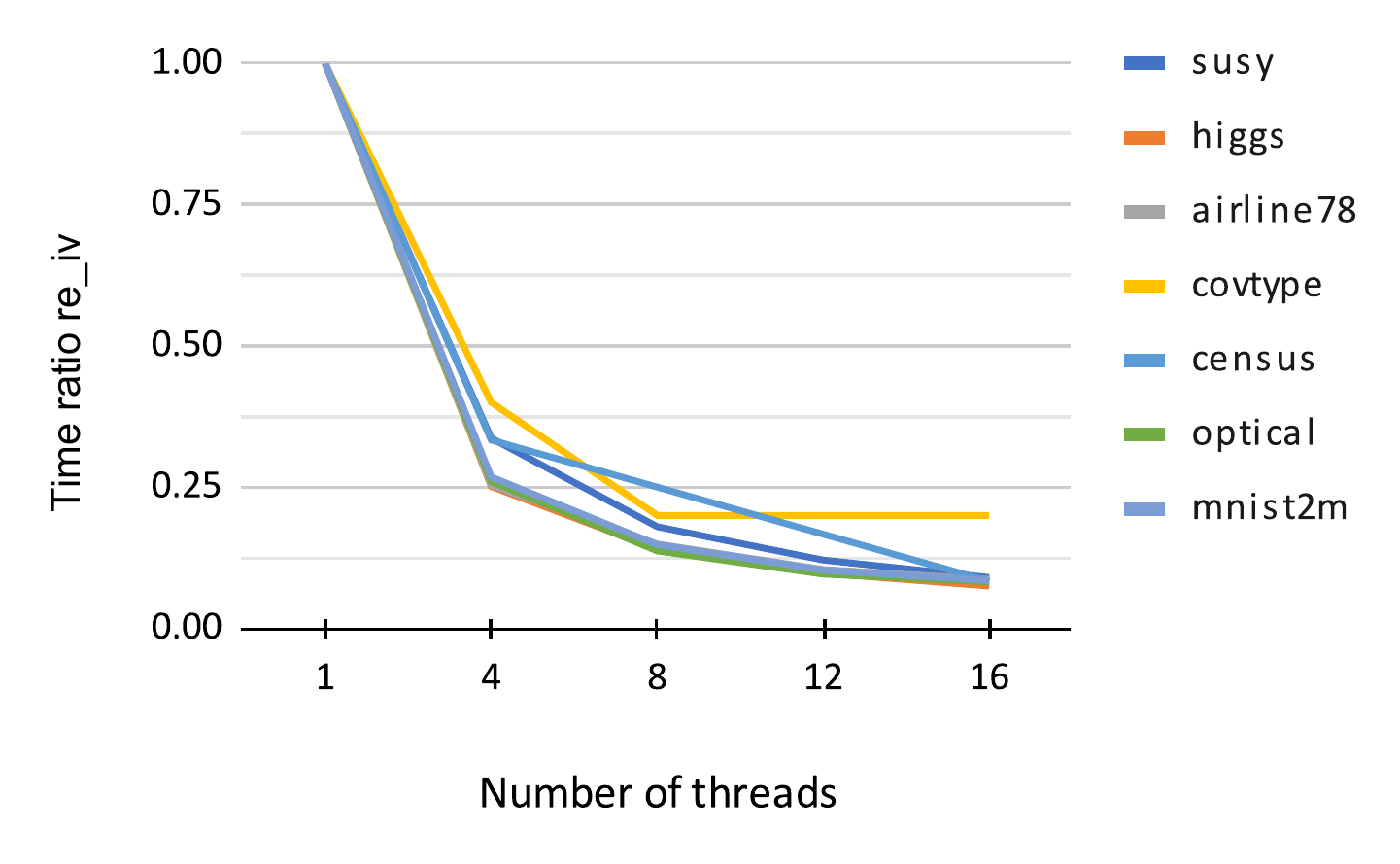}
\end{subfigure}
\caption{Peak memory usage (up) and running time (bottom) of the multithreaded version of the matrix multiplication algorithm using the \reansiv (left) and \reiv (right) compressors. The $Y$-axis reports the ratio between time and space of the multithreaded version of \reansiv or \reiv vs the single thread version of the same algorithm.}
\label{fig:mth}
\end{figure*}

For the above iterative scheme we measured the {\em average time per iteration} and the {\em peak memory usage} over the whole computation using the Unix tool {\tt time}. The results are reported in Table~\ref{table:iterations} and Figure~\ref{fig:mth}. In addition to the single thread algorithms, we tested the algorithms using 4, 8, 12, and 16 threads by splitting the input matrix into a number of row-blocks equal to the number of threads, as described in Section~\ref{sec:parallel}. The first two columns in Table~\ref{table:iterations} report the peak memory usage and average iteration time for the single thread version of the \reiv and \reansiv algorithms (i.e. the version in which the input matrix is not partitioned and grammar-compressed as a single unit). 

\begin{table*}[t]
\caption{Peak memory usage and average time per iteration in seconds for the computation of 500 iterations of Eq.~\eqref{eq:CG}. The memory usage is expressed as the percentage of the size of the full uncompressed matrix.}\label{table:iterations}
%	\centering
	\setlength{\tabcolsep}{5pt} % Default value: 6pt
	\renewcommand{\arraystretch}{1.2} % Default value: 1
	\scalebox{0.8}{\begin{tabular}{l||r|r||r|r||r|r||r|r||r|r||r|r}
		\hline 
		& \multicolumn{2}{c||}{\reiv 1 thread}
		 & \multicolumn{2}{c||}{\reansiv 1 thread} & \multicolumn{2}{c||}{\csrv 16 threads} & 
		 \multicolumn{2}{c||}{\reint 16 threads}& 
		 \multicolumn{2}{c||}{\reiv 16 threads}  & \multicolumn{2}{c}{\reansiv  16 threads}  \\ \hline
matrix   & peak mem &time & peak mem &time & peak mem &time & peak mem& time & peak mem & time & peak mem & time\\\hline

Susy     & 76.15\% & 3.89 & 73.40\% &  4.88 & 80.66\% & 0.26 & 80.63\% & 0.27 & 77.45\% & 0.35 & 82.67\% & 0.45 \\
Higgs    & 50.30\% & 8.28 & 47.12\% & 11.03 & 54.12\% & 0.36 & 52.04\% & 0.42 & 47.01\% & 0.62 & 44.90\% & 0.74 \\
Airline78& 17.16\% & 2.88 & 15.40\% &  3.94 & 41.57\% & 0.17 & 24.72\% & 0.15 & 19.21\% & 0.25 & 19.28\% & 0.31 \\
Covtype  &  9.42\% & 0.05 & 10.16\% &  0.07 & 14.60\% & 0.01 & 13.09\% & 0.01 & 17.10\% & 0.01 & 17.29\% & 0.01 \\
Census   &  4.37\% & 0.12 &  4.11\% &  0.19 & 23.88\% & 0.05 &  6.70\% & 0.01 &  6.14\% & 0.01 &  8.03\% & 0.02 \\
Optical  & 39.83\% & 0.73 & 39.23\% &  1.08 & 51.70\% & 0.04 & 46.56\% & 0.04 & 45.00\% & 0.06 & 56.72\% & 0.09 \\
Mnist2m  &  7.33\% & 7.09 &  6.85\% &  9.87 & 12.83\% & 0.20 & 11.31\% & 0.42 &  8.19\% & 0.60 &  8.30\% & 0.78 \\
\hline
\end{tabular}}
\end{table*}

% ; column $\Delta$ reports the difference between peak usage and the compressed size reported in Table~\ref{table:dataset}

As expected for both algorithms the peak memory usage of the single thread version of \reiv and \reansiv is somewhat larger than the compressed size reported in Table~\ref{table:dataset}. Indeed, according to Theorems~\ref{theo:y=mx} and~\ref{theo:x=ym} in addition to the space for the input and output vectors our algorithms use as a working space an (uncompressed) array of $|\Ru|$ 8-byte doubles. However, the difference between peak memory usage and compressed matrix size is less than 7\% of the uncompressed matrix size, with the only exception of \higgs for which it is $\approx$ 9\%. Unfortunately, the time per iteration of the single thread version is disappointing especially for the larger matrices. Hence, we have investigated the use of multiple threads by partitioning the matrix into blocks of consecutive rows as mentioned above.

Figure~\ref{fig:mth} shows the increase of the peak memory usage and the decrease of the running time as the number of threads increases for the algorithms \reansiv and \reiv. We see that, with the exception of the most compressible inputs (i.e. \covtype and \census), the peak memory usage using 16 threads is always less than 1.5 times the peak memory usage of the single thread version. Notice also that for \higgs the space usage of the multithread versions of \reiv and \reansiv is smaller than for the single thread version: the reason is that this file is better compressed when split into distinct blocks (this usually happens when the blocks have little structure in common). Comparing the plots at the top of Figure~\ref{fig:mth} we see that for \reiv the memory overhead of using multiple threads grows more slowly than for \reansiv. As a result, although \reiv is a simpler and usually less effective compressor, it uses less space than \reansiv when working with 16 threads as shown by the last two columns of  Table~\ref{table:iterations}. 

From Figure~\ref{fig:mth} (bottom left) we see that for \reansiv using 4 threads the speedup is close to 100\%, when using 8 threads the speedup is still close to the optimal apart from \census and \susy. As expected, a larger number of threads only helps \reansiv with the largest inputs: for \higgs, \airlinexx and \mnistxm with 16 threads the speedup is still within 12.66 and 14.90. On the other hand, for \covtype, which is the smallest input, \reansiv does not achieve any improvement by going from 8 to 16 threads.  
For \reiv the speedup follows a similar trend (Figure~\ref{fig:mth} bottom right). We notice that for 4 and 8 threads the speedup is smaller than for \reansiv, however for 16 threads the speedup is larger than 11 for all inputs except \covtype.

Table~\ref{table:iterations} summarises the statistics for the iterative computation of~\eqref{eq:CG} with \csrv and our grammar compressors using 16 threads. The results show that even for a real-life multithread computation the overall space usage can be still a small fraction of the uncompressed size. Indeed, the peak memory usage of the grammar compressors is up to 3 times smaller than for \csrv (i.e. for \census) and for 4 inputs it is less than 20\% of the original uncompressed size.  
As we will discuss in Section~\ref{sec:CLA}, such impressive compression rates achieve at the same time an average time per iteration which is always smaller than the time achieved by the state-of-the-art tool CLA.

The combined analysis of the peak memory usage vs running time highlights some interesting points. Considering all the algorithms running with 16 threads we see that, not surprisingly, the simpler compressed representations usually lead to faster matrix multiplications. Among the grammar compressors, the fastest algorithm is \reint in which the string $\Cset$ and the rule set $\Ru$ are represented with 32 bit integers. The more sophisticated encoders \reiv and \reansiv achieve better compression but they are slower. This is in accordance with the theoretical results:  according to Theorems~\ref{theo:y=mx} and~\ref{theo:x=ym} the cost of matrix-vector multiplication is $\Oh(|\Cset|+|\Ru|)$ time; \reiv and \reansiv use compressed representations of $\Cset$ and $\Ru$: this reduces the peak memory space but not the number of arithmetic operations. Including also the \csrv representation in the comparison (see column 3 in Table \ref{table:iterations}), we notice that different input matrices can have very different behaviours. For \airlinexx for example \reint uses much less space than \csrv but shows only a modest improvement in running time. For \mnistxm \reint again shows a modest reduction in space but an increase in running time; the most sophisticated \reiv and \reansiv tools get more compression but they are significantly slower. Finally, for \census we have a four times reduction in space for \reint and \reiv with a five-fold improvement in running time. 

Considering that the users want the fastest algorithm than can be run in the available memory, the conclusion is that all compressors should be considered and indeed an interesting problem would be the design of a mechanism for selecting the best options given the user's constraints. 

Finally, we point out that there are avenues for improving our  algorithms. For example, in our tests we used the same compressor for each row block of the input matrix: we could use different compressors to compress different blocks, or use the CSRV representation for the blocks which are hard to compress (a similar idea, applied to blocks of columns, is used within CLA). Another avenue for improvement is the reordering of the elements of the array $S$ as discussed at the end of Section~\ref{sec:proposal}: some promising results in this direction are presented in the next section.

\section{Column reordering for grammar compression}
\label{sec:colreoder}

In this section we show how the reordering of the columns of the input matrix improves the performance of our grammar compressor. As we mentioned at the end of Section~\ref{sec:proposal}, reordering the columns is only one of the possible preprocessing operations that can be applied to the input matrix without affecting our multiplication algorithms. We start our analysis with this technique as it was already studied in the related area of table compression~\cite{BFG03,VO2007273}.

Grammar compression for the CSRV representation works by replacing pairs of symbols appearing adjacent and in multiple rows with a single nonterminal. Therefore, we would like to reorder the matrix columns so that {\em correlated} columns appear adjacent to each other. To this end we define the {\em similarity} between two columns by measuring the number of identical pairs they form (cf.~the formal definition in the next subsection). In a sense, this similarity score estimates the compressible fraction of every column pair. This models the compression performance of a tool like RePair when the two columns are placed one adjacent to the other in the final ordering. This conservative, yet simple idea, achieves an effective performance as proved by our experiments. After having defined a notion of similarity between pairs of columns, we use it with four novel column-reordering algorithms and measure their impact on the performance of our compressed matrix-vector multiplication algorithm.

\subsection{The column-column similarity matrix}
\label{sub:csm}

Given the input matrix $M\in \RsupMN$ we define the $m\times m$ column-column similarity matrix $\CSM$ as follows. For each pair of column indices $i$ and $j$,  with $1 \leq i \neq j \leq m$, we build the sequence of pairs
\begin{equation*}
 P_{ij}=  \langle M[1][i], M[1][j]\rangle, \langle M[2][i], M[2][j]\rangle 
    \;\ldots\; \langle M[n][i], M[n][j]\rangle   
\end{equation*}
and we define $RP^{NZ}_{ij}$ as the number of repetitions of pairs of non-zero elements in the sequence $P_{ij}$ (note we only consider pairs in which both elements are nonzeros). For example for the matrix of Fig.~\ref{fig:vc} it holds $RP^{NZ}_{12}=2$ because $P_{12}$ contains only one non-zero pair, i.e. $\langle 1.2, 3.4\rangle$, which has two repetitions; and $RP^{NZ}_{13}= 1$ because $P_{24}$ contains two non-zero pairs, i.e. $\langle 1.2, 5.6\rangle$ and $\langle 1.2, 2.3\rangle$, but only one repetition of $\langle 1.2, 2.3\rangle$.

So, we define the similarity between columns $i$ and $j$ as the ratio
$$
    \CSM[j][i] = \CSM[i][j] = \frac{RP^{NZ}_{ij}}{n}.
$$
From the previous example we have $\CSM[1][2]= 2/6 = 0.\bar{3}$, and $\CSM[1][3] = 1/6 = 0.1\bar{6}$.

The computation of $\CSM[i][j]$ can be done in $O(n)$ expected time by inserting each pair in a hash table, thus taking $O(m^2 n)$ time over all column pairs. An alternative (computationally slower) procedure, which takes $O(m^2 n\log n)$ time overall, consists in collecting all pairs and sorting them in order to easily count duplicates. The sorting-based approach turned out to be very fast in practice and was the method of choice for our experiments.

The storage of $\CSM$ takes $\Theta(m^2)$ words if we use a full-sized representation. We also experimented with two heuristics as for reducing that space bound to $O(mk)$, where $k$ is a user-defined sparsity parameter. The first heuristic consists of building a sparse $\CSM$ matrix, called {\em locally-pruned} column-column similarity matrix $\CSM_P$, in which we maintain only the $k$ greatest column-column similarity scores for each column. The second heuristic builds a globally-pruned column-column similarity matrix $\CSM_P$ by keeping the top-$(mk)$ similarity scores among all the entries of $\CSM$. The space complexity is still $O(mk)$, but now the pruning is performed {\em globally} over all entries of the original matrix.

\subsection{Column-reordering approaches}
\label{sub:reorder}

Once we have computed the column-similarity matrix $\CSM$ either in its full or sparse version, we leverage it to find a column reordering that helps grammar compression. We investigated four different column-reordering algorithms, which work on the weighted graph $G$ whose adjacency matrix is either $\CSM$ (and thus consists of $\Theta(m^2)$ edges) or $\CSM_P$ (and thus consists of $\Theta(mk)$ edges). They are described below.

The {\bf Lin-Kernighan heuristic (LKH)} is a well-known approach to the solution of the Travelling Salesman Problem (TSP). Even though the algorithm is approximate, the implementation in \cite{Helsgaun:2000,Helsgaun:2009} computes the best known solution for a series of large-scale instances with unknown optima.
	We model column reordering as an instance of a (symmetric) TSP problem stated on the graph $G$ above. Each of the $m$ columns in the original matrix $M$ corresponds to a different city in the TSP; the distance between pairs of cities (columns) is given by the corresponding entry in the matrices $\CSM$ or $\CSM_P$ (negated, since the TSP is a minimisation problem and we want to maximise total similarity). The TSP solution will specify an ordering of $M$'s columns. For the experiments, we use the {ANSI C} implementation of \LKH available at: \url{http://webhotel4.ruc.dk/~keld/research/LKH/} (version 2.0.9).
	
	The {\bf PathCover} approach models column reordering as the problem of finding a set of maximum weighted paths in the weighted undirected graph $G$ mentioned above, with the additional requirements that these paths ``cover'' all of its nodes and they are disjoint. We introduce this approach starting from the consideration that TSP is a hard problem, but we do not necessarily need to impose its strong constraint of forming a single Hamiltonian path. Indeed it may be a good idea to concentrate our algorithmic effort upon the subset of columns that are more compressible \cite{Shi_primary_set}, leaving aside those columns that do not exhibit significant redundancies and correlations.
	{\PathCover} turns out to be a much faster approach because it returns a set of partial reorderings (induced by the found paths), which may be eventually concatenated together to form a full reordering. This approach is a reminiscence of the {\em single linkage} algorithm used in hierarchical clustering~\cite[Ch.~17]{MRS:2008}. 
	We implement {\PathCover} using a variant of the Kruskal's algorithm for Minimum Spanning Trees~\cite{CLRS:2009}. Our algorithm scans $G$'s edges by decreasing weights, and adds edges to the solution only if they form disjoint paths. Our implementation of {\PathCover} is written in Python, yet it is very fast in practice.

	{\bf PathCover+} is a variant of {\PathCover} in which the column-column similarity matrix is dynamically updated as follows. Let $(u_{r-1}, u_r)$ be the heaviest edge selected by the {\PathCover} algorithm, and assume that it extends a covering path to form ${\mathcal P} = (u_1, ..., u_{r-1}, u_r)$. Then, for each node $v$ adjacent to some node $u_j \in {\mathcal P}$, we recompute the new weight $w(v,u_j)$ as the minimum among the weights from $v$ to any node in ${\mathcal P}$. Thus, the weighting corresponds to coalescing the path ${\mathcal P}$ into a macro-node and making the link from $v$ to ${\mathcal P}$ as the minimum weighted edge from $v$ to any node $u \in {\mathcal P}$.  We implemented \PathCoverP in Python  following a procedure similar to Sybein's MST algorithm \cite{Mehlhorn_Sanders_book_chap11}.
	
% This approach is a reminiscence of the {\em complete linkage} algorithm used for hierarchical clustering~\cite[Ch.~17]{MRS:2008}.

	The {\bf Maximum Weighted Matching (MWM)} approach consists in computing a weighted matching ${\mathcal M}$ of the graph $G$. By definition, ${\mathcal M}$ is a subgraph of $G$ such that no two edges share common vertices and the sum of the edge weights is maximum among all possible matchings in $G$. The best exact \MWM algorithm exhibits $\Theta(m^3)$ time complexity \cite{Gabow:1976}.
	For our column-reordering purpose, we generate a bipartite graph $B_G$ with $2m$ nodes and ${m \choose 2}$ edges weighted according to the column-column similarity entries. More in detail, for each column pair $i$, $j$, with $i<j$, we insert an edge in $B_G$ that connects the $i$-th node to the $j$-th node and assign to it weight $\CSM[i][j]$ or $\CSM_P[i][j]$. Choosing that edge corresponds to assuming that the $i$-th column precedes the $j$-th column in the column reordering. 
	After the \MWM is computed, we use this predecessor-successor relation to determine the final column reordering. Notice that cycles cannot be induced because we assumed that edges $(i,j)$ are oriented, namely $i<j$. If the matching size $|{\mathcal M}|$ is lower than the number of columns $m$ in $M$, then \MWM does not induce a single column-reordering sequence, but rather a set of shorter disjoint column-reordering sequences: we thus concatenate these partial reordering sequences in an arbitrary order to form a full column reordering.
	We implemented \MWM in {\tt C++} using the Boost library (\url{https://www.boost.org}).

\subsection{Experimenting with column reordering}
\label{sec:cr_exp}

We conducted a set of experiments using the matrices reported in Table~\ref{table:dataset} to analyse the time and space performance of the column-reordering approaches described above. After applying the column-reordering algorithm we compressed the reordered matrix using \reansiv from Section~\ref{sec:implementation}.
We report the results only for the methods \LKH, \PathCover, and \MWM, since the application of the \PathCoverP method always resulted in worse compression performance.

The three column-reordering algorithms exhibit qui\-te different time performance. \PathCover is the fastest algorithm, whereas \MWM and \LKH especially are the most time-consuming ones.
On the {\mnistxm} matrix (which has the largest number of columns) \PathCover takes roughly {one} second, while \MWM takes about $20$ seconds, and \LKH takes half an hour to compute its result. The running time of \LKH slightly varies with the setting of \LKH heuristic (faster solutions correspond to worse results), but in any case \LKH stays orders of magnitude slower than the other approaches.

In terms of space performance, we found that the locally-pruned version of the \CSM matrix usually performs better than the full matrix or the globally pruned matrix.   Table~\ref{tab:lp_original} reports the compression achieved by this approach for the three reordering algorithms and for three different values of the sparsity parameter~$k$. The compression ratios are relative to the size of the uncompressed matrix and therefore can be compared against those in Table~\ref{table:dataset}.
We see from Table~\ref{tab:lp_original} that for \susy the three reordering algorithms exhibit the same performance, while for the other six matrices, \PathCover is superior over three of them, while \MWM is the winner for the other three.  \LKH is often very close to the best compression, but since it is the most computationally expensive algorithm, we conclude that it is overall not a competitive solution.

\begin{table}[t]
\caption{Compression achieved by our column-reordering algorithms, with the {locally-pruned} $\CSM$ matrix, followed by the algorithm \reansiv. The percentages refer to the size of the uncompressed matrix and therefore can be compared with those in Table~\ref{table:dataset}. Green (red) values highlight the best (worse) compression for each matrix.\label{tab:lp_original}} 
	%\centering
	\scalebox{.8}{\begin{tabular}{l|l||c|c|c}
		\hline
\multicolumn{2}{l||}{matrix}  & \LKH & \PathCover & \MWM\\ \hline
		&
		$k=4$ &
		{66.57\%} &
		{66.57\%} &
		{66.57\%} \\ \cline{3-5} 
		&
		$k=8$ &
		{66.57\%} &
		{66.57\%} &
		{66.57\%} \\ \cline{3-5} 
		\multirow{-3}{*}{Susy} &
		$k=16$ &
		{66.57\%} &
		{66.57\%} &
		{66.57\%} \\ \hline \hline
		&
		$k=4$ &
		{38.03\%} &
		38.00\% &
		{37.99\%} \\ \cline{3-5} 
		&
		$k=8$ &
		37.92\% &
		38.00\% &
		{37.98\%} \\ \cline{3-5} 
		\multirow{-3}{*}{Higgs} &
		$k=16$ &
		{38.02\%} &
		\worst{38.04\%} &
		\best{37.92\%} \\ \hline \hline
		&
		$k=4$ &
		{9.63\%} &
		{9.21\%} &
		\worst{10.17\%} \\ \cline{3-5} 
		&
		$k=8$ &
		{8.65\%} &
		{9.52\%} &
		\best{8.32\%} \\ \cline{3-5} 
		\multirow{-3}{*}{Airline78} &
		$k=16$ &
		{9.43\%} &
		{8.34\%} &
		{9.63\%} \\ \hline \hline 
		&
		$k=4$ &
		3.74\% &
		{3.30\%} &
		\worst{4.19\%} \\ \cline{3-5} 
		&
		$k=8$ &
		3.51\% &
		\best{3.24\%} &
		3.72\% \\ \cline{3-5} 
		\multirow{-3}{*}{Covtype} &
		$k=16$ &
		3.25\% &
		3.26\% &
		{3.72\%} \\ \hline \hline
		&
		$k=4$ &
		1.37\% &
		{1.39\%} &
		1.37\% \\ \cline{3-5} 
		&
		$k=8$ &
		1.33\% &
		1.37\% &
		\worst{1.41\%} \\ \cline{3-5} 
		\multirow{-3}{*}{Census} &
		$k=16$ &
		{1.31\%} &
		\best{1.30\%} &
		1.39\% \\ \hline \hline
		
		&
		$k=4$ &
		{33.23\%} &
		\best{32.60\%} &
		{33.19\%} \\ \cline{3-5} 
		&
		$k=8$ &
		{32.68\%} &
		{33.03\%} &
		\worst{33.26\%} \\ \cline{3-5} 
		\multirow{-3}{*}{Optical} &
		$k=16$ &
		{33.22\%} &
		{32.89\%} &
		{32.95\%} \\ \hline \hline
		&
		$k=4$ &
		5.29\% &
		5.31\% &
		\worst{5.32\%} \\ \cline{3-5} 
		&
		$k=8$ &
		{5.29\%} &
		5.31\% &
		\best{5.29\%} \\ \cline{3-5} 
		\multirow{-3}{*}{Mnist2m} &
		$k=16$ &
		5.30\% &
		{5.31\%} &
		5.30\% \\ \hline %\hline
		
	\end{tabular}}
\end{table}

\newcommand\rereansiv{\mathsf{reordered\_re\_ans}}

\begin{table*}[t]
\caption{Performance of \reiv and \reansiv applied to the optimally blockwise reordered matrix. Size is expressed in percentage with respect to the full uncompressed matrix. The last three columns report the performance of CLA. For CLA, peak memory and time include the resources used for matrix compression; see discussion in Section~\ref{sec:CLA}.}\label{table:permvscla}
	%\centering
	\setlength{\tabcolsep}{6pt} % Default value: 6pt
	\renewcommand{\arraystretch}{1.2} % Default value: 1
\scalebox{.8}{\begin{tabular}{l||r|r|r||r|r|r||r|r|r}
\hline
%\multicolumn{2}{c}{CLA} \\ \hline
         & \multicolumn{3}{c||}{\reiv 16 threads}  &
         \multicolumn{3}{c||}{\reansiv 16 threads}  &
           \multicolumn{3}{c}{CLA multithread} \\ \hline
matrix   & 
\multicolumn{1}{c|}{size} & \multicolumn{1}{c|}{peak mem}  & \multicolumn{1}{c||}{time} &
\multicolumn{1}{c|}{size} & \multicolumn{1}{c|}{peak mem}  & \multicolumn{1}{c||}{time} &
\multicolumn{1}{c|}{size} & \multicolumn{1}{c|}{peak mem}  & \multicolumn{1}{c}{time} \\ \hline

Susy     & 68.99\% & 77.53\% & 0.35 & 65.99\% & 82.77\% & 0.45 &76.14\% & ---       & ---  \\
Higgs    & 41.63\% & 46.68\% & 0.58 & 37.44\% & 44.63\% & 0.71 &32.74\% &12354.93\% & 2.09 \\
Airline78&  9.35\% & 16.06\% & 0.17 &  8.13\% & 16.43\% & 0.23 &12.34\% & 2279.29\% & 1.17 \\
Covtype  &  4.78\% & 16.25\% & 0.01 &  4.17\% & 16.11\% & 0.01 & 4.55\% & 1008.50\% & 0.05 \\
Census   &  2.00\% &  5.70\% & 0.01 &  1.55\% &  7.25\% & 0.02 & 3.77\% &  354.00\% & 0.16 \\
Optical  & 36.05\% & 44.50\% & 0.06 & 34.93\% & 56.39\% & 0.09 &40.44\% & 3779.65\% & 0.20 \\
Mnist2m  &  6.24\% &  8.19\% & 0.64 &  5.88\% &  8.30\% & 0.82 & 6.22\% &  382.52\% & 1.98 \\

\hline
\end{tabular}}
\end{table*}

To measure the effectiveness of the reordering techniques on the matrix multiplication operations we performed the following experiment. 
We partitioned each input matrix in 16 blocks of rows as described in Section~\ref{sec:parallel}. Then, we applied to each block 
\PathCover and \MWM (in both cases with sparsity parameter $k=16$) followed by \reansiv and selected the reordering algorithm yielding the best overall compression. In other words, for each matrix we select the best reordering algorithm between \PathCover and \MWM for its blocks, and compress the single blocks individually and independently by \reansiv, each one possibly with a different permutation.\footnote{As we observed at the end of Section \ref{sec:lmultiplication}, we do not need to store the column permutation because every pair in $S$ keeps the original column of each element.} With such reordered matrix, compressed with \reansiv, we performed our benchmark computation~(\ref{eq:CG}) and recorded the peak memory usage and the average time per iteration. We repeated the same procedure for the algorithm \reiv and reported all the measurements in Table~\ref{table:permvscla}. Comparing those results with those in Table~\ref{table:iterations} we see that reordering helps to reduce the peak space and, to a lesser degree the average running time. 

\begin{figure}[h]
\includegraphics[width=0.80\linewidth]{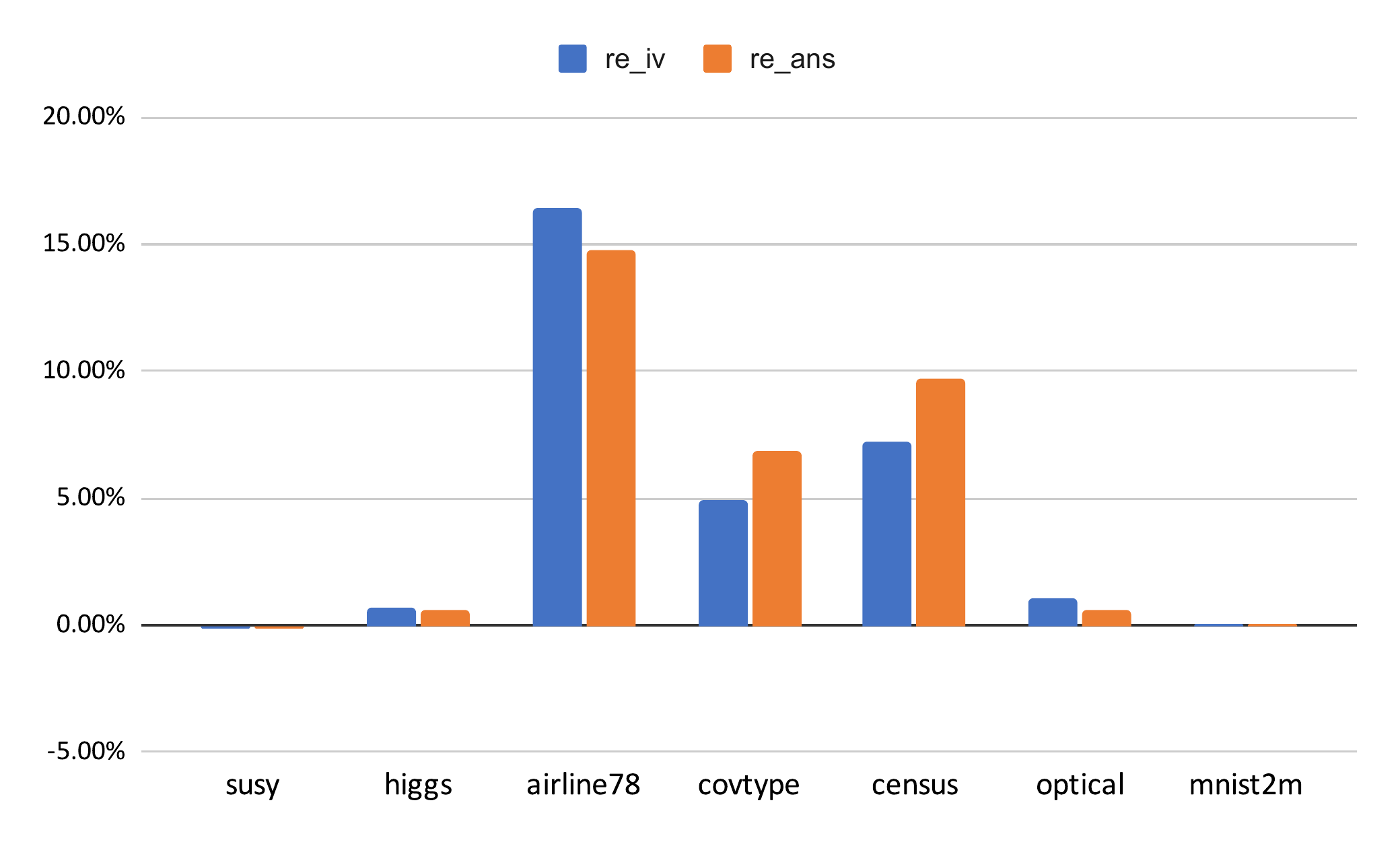}
\caption{Percentage (relative) improvements in terms of the peak memory usage for the reordered matrices as resulting from the data reported in Table~\ref{table:permvscla} for the compressors based on \reiv and \reansiv.}
\label{fig:peak}
\end{figure}

Although the benefits of reordering might appear small in absolute terms they are significant with respect to the size of the compressed matrix. To see this,  Figure~\ref{fig:peak} shows, for the two algorithms and for each input matrix, the percentage improvements in the peak memory usage, computed as $(p_o - p_r)/p_o$ where $p_o$ and $p_r$ are respectively the peak memory usage for the original and for the reordered matrix. We see that there is a significant memory-usage reduction for three inputs \airlinexx, \covtype, and \census which are compressed by \reiv and \reansiv to less that 20\% of their original size. Remarkably, for \airlinexx such memory reduction translates to a 25\% reduction in the average running time. Note also that sometimes reordering does not help: for \mnistxm reordering does not change the peak memory usage but induces a small (5\%) increase in the running time for both algorithms; and for \susy, the reordering slightly increases the peak memory usage, with no significant changes in the running time. 

\ignore{Considering that \PathCover and \MWM  run in less than one minute, the use of a reordering step appears to be worthwhile and encourages the investigation of more sophisticated matrix preprocessing techniques. As we have already pointed out, our multiplication algorithms do support without any modification the application of different column {permutations} to different groups of rows. }

\subsection{Comparison with CLA}\label{sec:CLA}

As mentioned in the Introduction, our research takes inspiration from the results by Elgohary et al.~\cite{elgohary2019compressed,elgohary19matrix} on Compressed Linear Algebra (CLA) where they introduced the idea of specialised compression techniques supporting matrix operations without decompression. 
Thus, it is of interest to compare CLA compression against our grammar compressors in terms of space usage and speed. 
Unfortunately, the comparison with CLA faces some technical hurdles: we implemented our tools in small self-contained C/C++ programs, while CLA is available inside Apache SystemDS~\cite{SystemDS}, a complete Machine Learning system written in Java and designed to run possibly on top of Apache Spark. Inside SystemDS, algorithms are expressed in a high-level language with an R-like syntax: such scripts are parsed and analysed before the actual computation starts. 
In addition, SystemDS does not store the compressed data on disk: matrices are compressed from scratch at every execution, and since the compression algorithm has a randomised component the compressed representation can change from one execution to the next one. Within such a complex system there is not an accurate way\footnote{We are currently trying to overcome this issue with the help of CLA developers.} to determine the actual memory footprint of the sole matrix multiplication algorithm.

% that is, a meaningful space cost measure to be compared against the ``peak mem'' column in Table~\ref{table:iterations}.   

With these limitations in mind, since we wanted to compare our approach against CLA, we nevertheless set up a rough comparison which we believe offers an interesting viewpoint on their relative performance. We wrote a script executing 500 iterations of Eq.~\eqref{eq:CG} and we ran it on SystemDS over the matrices of our data set. The results of these experiments are reported in the last three columns of Table~\ref{table:permvscla}. Column ``size'' reports the size of the 
CLA's compressed representation of the input matrix, obtained from the system log file.  Such size is expressed as percentage with respect to the uncompressed dense representation hence it can directly compared with the other ``size'' columns of Table~\ref{table:permvscla} and with the sizes in the last six columns of Table~\ref{table:dataset}. We see that in terms of compression CLA is less effective than \reansiv with the only exception of \higgs. Compared to \reiv CLA is clearly superior for \higgs, marginally superior for \covtype and \mnistxm, and less effective for all the other inputs. 

The last two columns of Table~\ref{table:permvscla} show the peak memory usage and average running time as reported by the Unix tool {\tt time}. As we mentioned above, such measurements include the compression of the input matrix which is done from scratch at each invocation. While this is not a serious issue for the running time, since the logs show that compression is fast and its contribution is averaged over 500 iterations, the peak memory usage is likely to be achieved  in the compression phase so the reported figures are just an upper bound to the memory usage during the multiplication phase. Note that for the input \susy CLA was unable to complete the multiplication phase because of a Java runtime exception. 

The comparison of the running times shows that \reansiv was always at least twice faster than CLA and \reiv was always at least three times faster than CLA. Note that these results were obtained using 16 threads for \reansiv and \reiv, while CLA is designed to use all the available threads (the testing machine supports up to 80 independent threads). In addition {CLA uses} a preliminary planning phase in which rows and columns are permuted and grouped to help compression and make the matrix operations more cache-conscious. By comparison, our strategy of partitioning the input matrix into a fixed number of row blocks and possibly reordering the columns of each block is certainly much simpler.

Summing up, although rough, our comparison allows us to draw some important conclusions: (1) grammar compressors are indeed able to achieve a better space reduction than CLA, which is not surprising given that they are provably able to compress their input up to the $k$-th order statistical entropy, and (2) the theoretical results ensuring that the number of operations is bounded by the size of the compressed matrix translate to algorithms that are also faster in practice to compute matrix operations.

\section{Conclusions and Future Work}

We have presented a grammar-based lossless compression scheme for real-valued matrices. The use of grammar compressors guarantees that the size of the compressed matrix is proportional to the $k$-th order statistical entropy of the Compressed Sparse Row/Value matrix representation. We have shown that we can compute the left and the right matrix-vector multiplications in time and space linear in the size of the compressed matrix representation. 

%We have implemented and tested three variants of a grammar-based compression algorithm, showing different time/space tradeoffs. 

These remarkable properties of our approach open the related problem of reordering the matrix elements in order to maximise compression. This requires discovering and exploiting the usually hidden dependencies between the elements in ML matrices. As a first step in this direction we have introduced and tested four column-reordering algorithms based on a new column-similarity score, which takes into account the subsequent application of a grammar-based compressor. Experimental results have shown that with a modest preprocessing time, our column reordering can yield a further reduction in the storage space of the compressed matrix and a speed-up in the matrix-vector multiplication time. All these achievements improve the results got by the state-of-the-art Compressed Linear Algebra (CLA) approach \cite{elgohary19matrix, elgohary2019compressed}.

As a future work, we plan to investigate how much row permutation and co-clustering techniques~\cite{jmlr/BanerjeeDGMM07,kdd/Dhillon01,HanL16} can contribute to achieve even better compression ratios. It would be of interest also to adapt and test our matrix-compression scheme in the context of columnar DBs, which feature multiple data types, such as strings, integers, categorical data, etc.. Finally, Web and Social graphs offer another relevant opportunity for the application of our new compression schemes.

%%%%%%%%%%%%%%%%%%%%%%%%%%%%%%
% BIBLIO
%%%%%%%%%%%%%%%%%%%%%%%%%%%%%%

%\clearpage

\bibliographystyle{ACM-Reference-Format}
\bibliography{mrepair}


\begin{thebibliography}{34}

%%% the Web address.

\ifx \showCODEN    \undefined \def \showCODEN     #1{\unskip}     \fi
\ifx \showDOI      \undefined \def \showDOI       #1{#1}\fi
\ifx \showISBNx    \undefined \def \showISBNx     #1{\unskip}     \fi
\ifx \showISBNxiii \undefined \def \showISBNxiii  #1{\unskip}     \fi
\ifx \showISSN     \undefined \def \showISSN      #1{\unskip}     \fi
\ifx \showLCCN     \undefined \def \showLCCN      #1{\unskip}     \fi
\ifx \shownote     \undefined \def \shownote      #1{#1}          \fi
\ifx \showarticletitle \undefined \def \showarticletitle #1{#1}   \fi
\ifx \showURL      \undefined \def \showURL       {\relax}        \fi
% invisible to TeX
\providecommand\bibfield[2]{#2}
\providecommand\bibinfo[2]{#2}
\providecommand\natexlab[1]{#1}
\providecommand\showeprint[2][]{arXiv:#2}

\bibitem[\protect\citeauthoryear{Apostolico, Cunial, and Kaul}{Apostolico
  et~al\mbox{.}}{2008}]%
        {apostolico08}
\bibfield{author}{\bibinfo{person}{Alberto Apostolico}, \bibinfo{person}{Fabio
  Cunial}, {and} \bibinfo{person}{Vineith Kaul}.}
  \bibinfo{year}{2008}\natexlab{}.
\newblock \showarticletitle{Table Compression by Record Intersections}. In
  \bibinfo{booktitle}{\emph{2008 Data Compression Conference {(DCC} 2008),
  25-27 March 2008, Snowbird, UT, {USA}}}. \bibinfo{publisher}{{IEEE} Computer
  Society}, \bibinfo{pages}{13--22}.
\newblock


\bibitem[\protect\citeauthoryear{ASA, Statistical Computing \& Graphics
  Sections}{ASA, Statistical Computing \& Graphics Sections}{2021}]%
        {Airline}
ASA, Statistical Computing \& Graphics Sections
  \bibinfo{year}{2021}\natexlab{}.
\newblock \bibinfo{title}{{Data Expo 2009 -- Airline on-time performance}}.
\newblock
  \bibinfo{howpublished}{\url{https://community.amstat.org/jointscsg-section/dataexpo/dataexpo2009}}.
\newblock
\newblock
\shownote{[Online; accessed 21-Sep-2021].}


\bibitem[\protect\citeauthoryear{Banerjee, Dhillon, Ghosh, Merugu, and
  Modha}{Banerjee et~al\mbox{.}}{2007}]%
        {jmlr/BanerjeeDGMM07}
\bibfield{author}{\bibinfo{person}{Arindam Banerjee},
  \bibinfo{person}{Inderjit~S. Dhillon}, \bibinfo{person}{Joydeep Ghosh},
  \bibinfo{person}{Srujana Merugu}, {and} \bibinfo{person}{Dharmendra~S.
  Modha}.} \bibinfo{year}{2007}\natexlab{}.
\newblock \showarticletitle{A Generalized Maximum Entropy Approach to {Bregman}
  Co-clustering and Matrix Approximation}.
\newblock \bibinfo{journal}{\emph{J. Mach. Learn. Res.}}  \bibinfo{volume}{8}
  (\bibinfo{year}{2007}), \bibinfo{pages}{1919--1986}.
\newblock


\bibitem[\protect\citeauthoryear{Bottou}{Bottou}{2007}]%
        {MNIST}
\bibfield{author}{\bibinfo{person}{Leon Bottou}.}
  \bibinfo{year}{2007}\natexlab{}.
\newblock \bibinfo{title}{The infinite {MNIST} dataset}.
\newblock
  \bibinfo{howpublished}{\url{https://leon.bottou.org/projects/infimnist}}.
\newblock
\urldef\tempurl%
\url{https://leon.bottou.org/projects/infimnist}
\showURL{%
\tempurl}
\newblock
\shownote{[Online; accessed 21-Sep-2021].}


\bibitem[\protect\citeauthoryear{Buchsbaum, Fowler, and Giancarlo}{Buchsbaum
  et~al\mbox{.}}{2003}]%
        {BFG03}
\bibfield{author}{\bibinfo{person}{Alan~L. Buchsbaum},
  \bibinfo{person}{Glenn~S. Fowler}, {and} \bibinfo{person}{Raffaele
  Giancarlo}.} \bibinfo{year}{2003}\natexlab{}.
\newblock \showarticletitle{Improving Table Compression with Combinatorial
  Optimization}.
\newblock \bibinfo{journal}{\emph{J. ACM}}  \bibinfo{volume}{50}
  (\bibinfo{year}{2003}), \bibinfo{pages}{825--851}.
\newblock


\bibitem[\protect\citeauthoryear{Chakrabarti, Papadimitriou, Modha, and
  Faloutsos}{Chakrabarti et~al\mbox{.}}{2004}]%
        {ChakrabartiPMF04}
\bibfield{author}{\bibinfo{person}{Deepayan Chakrabarti},
  \bibinfo{person}{Spiros Papadimitriou}, \bibinfo{person}{Dharmendra~S.
  Modha}, {and} \bibinfo{person}{Christos Faloutsos}.}
  \bibinfo{year}{2004}\natexlab{}.
\newblock \showarticletitle{Fully automatic cross-associations}. In
  \bibinfo{booktitle}{\emph{Proceedings of the Tenth {ACM} {SIGKDD}
  International Conference on Knowledge Discovery and Data Mining, Seattle,
  Washington, USA, August 22-25, 2004}}. \bibinfo{publisher}{{ACM}},
  \bibinfo{pages}{79--88}.
\newblock


\bibitem[\protect\citeauthoryear{Charikar, Lehman, Liu, Panigrahy, Prabhakaran,
  Sahai, and Shelat}{Charikar et~al\mbox{.}}{2005}]%
        {tit/CharikarLLPPSS05}
\bibfield{author}{\bibinfo{person}{Moses Charikar}, \bibinfo{person}{Eric
  Lehman}, \bibinfo{person}{Ding Liu}, \bibinfo{person}{Rina Panigrahy},
  \bibinfo{person}{Manoj Prabhakaran}, \bibinfo{person}{Amit Sahai}, {and}
  \bibinfo{person}{Abhi Shelat}.} \bibinfo{year}{2005}\natexlab{}.
\newblock \showarticletitle{The smallest grammar problem}.
\newblock \bibinfo{journal}{\emph{{IEEE} Trans. Inf. Theory}}
  \bibinfo{volume}{51}, \bibinfo{number}{7} (\bibinfo{year}{2005}),
  \bibinfo{pages}{2554--2576}.
\newblock


\bibitem[\protect\citeauthoryear{Cormen, Leiserson, Rivest, and Stein}{Cormen
  et~al\mbox{.}}{2009}]%
        {CLRS:2009}
\bibfield{author}{\bibinfo{person}{Thomas~H. Cormen},
  \bibinfo{person}{Charles~E. Leiserson}, \bibinfo{person}{Ronald~L. Rivest},
  {and} \bibinfo{person}{Clifford Stein}.} \bibinfo{year}{2009}\natexlab{}.
\newblock \bibinfo{booktitle}{\emph{Introduction to Algorithms, 3rd Edition}}.
\newblock \bibinfo{publisher}{{MIT} Press}.
\newblock


\bibitem[\protect\citeauthoryear{Dhillon}{Dhillon}{2001}]%
        {kdd/Dhillon01}
\bibfield{author}{\bibinfo{person}{Inderjit~S. Dhillon}.}
  \bibinfo{year}{2001}\natexlab{}.
\newblock \showarticletitle{Co-clustering documents and words using bipartite
  spectral graph partitioning}. In \bibinfo{booktitle}{\emph{{KDD}}}.
  \bibinfo{publisher}{{ACM}}, \bibinfo{pages}{269--274}.
\newblock


\bibitem[\protect\citeauthoryear{Dua and Graff}{Dua and Graff}{2017}]%
        {UCI}
\bibfield{author}{\bibinfo{person}{Dheeru Dua} {and} \bibinfo{person}{Casey
  Graff}.} \bibinfo{year}{2017}\natexlab{}.
\newblock \bibinfo{title}{{UCI} Machine Learning Repository}.
\newblock \bibinfo{howpublished}{\url{https://archive.ics.uci.edu/ml}}.
\newblock
\urldef\tempurl%
\url{http://archive.ics.uci.edu/ml}
\showURL{%
\tempurl}
\newblock
\shownote{[Online; accessed 21-Sep-2021].}


\bibitem[\protect\citeauthoryear{Elgohary, Boehm, Haas, Reiss, and
  Reinwald}{Elgohary et~al\mbox{.}}{2018}]%
        {elgohary2019compressed}
\bibfield{author}{\bibinfo{person}{Ahmed Elgohary}, \bibinfo{person}{Matthias
  Boehm}, \bibinfo{person}{Peter~J. Haas}, \bibinfo{person}{Frederick~R.
  Reiss}, {and} \bibinfo{person}{Berthold Reinwald}.}
  \bibinfo{year}{2018}\natexlab{}.
\newblock \showarticletitle{Compressed linear algebra for large-scale machine
  learning}.
\newblock \bibinfo{journal}{\emph{{VLDB} J.}} \bibinfo{volume}{27},
  \bibinfo{number}{5} (\bibinfo{year}{2018}), \bibinfo{pages}{719--744}.
\newblock


\bibitem[\protect\citeauthoryear{Elgohary, Boehm, Haas, Reiss, and
  Reinwald}{Elgohary et~al\mbox{.}}{2019}]%
        {elgohary19matrix}
\bibfield{author}{\bibinfo{person}{Ahmed Elgohary}, \bibinfo{person}{Matthias
  Boehm}, \bibinfo{person}{Peter~J. Haas}, \bibinfo{person}{Frederick~R.
  Reiss}, {and} \bibinfo{person}{Berthold Reinwald}.}
  \bibinfo{year}{2019}\natexlab{}.
\newblock \showarticletitle{Compressed linear algebra for declarative
  large-scale machine learning}.
\newblock \bibinfo{journal}{\emph{Commun. {ACM}}} \bibinfo{volume}{62},
  \bibinfo{number}{5} (\bibinfo{year}{2019}), \bibinfo{pages}{83--91}.
\newblock
\urldef\tempurl%
\url{https://doi.org/10.1145/3318221}
\showDOI{\tempurl}


\bibitem[\protect\citeauthoryear{Francisco, Gagie, Ladra, and
  Navarro}{Francisco et~al\mbox{.}}{2018}]%
        {francisco2018exploiting}
\bibfield{author}{\bibinfo{person}{Alexandre Francisco},
  \bibinfo{person}{Travis Gagie}, \bibinfo{person}{Susana Ladra}, {and}
  \bibinfo{person}{Gonzalo Navarro}.} \bibinfo{year}{2018}\natexlab{}.
\newblock \showarticletitle{Exploiting computation-friendly graph compression
  methods for adjacency-matrix multiplication}. In
  \bibinfo{booktitle}{\emph{2018 Data Compression Conference}}.
  \bibinfo{publisher}{IEEE Computer Society Press}, \bibinfo{pages}{307--314}.
\newblock


\bibitem[\protect\citeauthoryear{Gabow}{Gabow}{1976}]%
        {Gabow:1976}
\bibfield{author}{\bibinfo{person}{Harold~N. Gabow}.}
  \bibinfo{year}{1976}\natexlab{}.
\newblock \showarticletitle{An Efficient Implementation of Edmonds' Algorithm
  for Maximum Matching on Graphs}.
\newblock \bibinfo{journal}{\emph{J. ACM}} \bibinfo{volume}{23},
  \bibinfo{number}{2} (\bibinfo{date}{April} \bibinfo{year}{1976}),
  \bibinfo{pages}{221–234}.
\newblock
\showISSN{0004-5411}
\urldef\tempurl%
\url{https://doi.org/10.1145/321941.321942}
\showDOI{\tempurl}


\bibitem[\protect\citeauthoryear{Gog, Beller, Moffat, and Petri}{Gog
  et~al\mbox{.}}{2014}]%
        {gbmp2014sea}
\bibfield{author}{\bibinfo{person}{Simon Gog}, \bibinfo{person}{Timo Beller},
  \bibinfo{person}{Alistair Moffat}, {and} \bibinfo{person}{Matthias Petri}.}
  \bibinfo{year}{2014}\natexlab{}.
\newblock \showarticletitle{From Theory to Practice: Plug and Play with
  Succinct Data Structures}. In \bibinfo{booktitle}{\emph{13th International
  Symposium on Experimental Algorithms, (SEA 2014)}}
  \emph{(\bibinfo{series}{LNCS})}, Vol.~\bibinfo{volume}{8504}.
  \bibinfo{publisher}{Springer}, \bibinfo{pages}{326--337}.
\newblock


\bibitem[\protect\citeauthoryear{Han and Li}{Han and Li}{2016}]%
        {HanL16}
\bibfield{author}{\bibinfo{person}{Bo Han} {and} \bibinfo{person}{Bolang Li}.}
  \bibinfo{year}{2016}\natexlab{}.
\newblock \showarticletitle{Lossless Compression of Data Tables in Mobile
  Devices by Using Co-clustering}.
\newblock \bibinfo{journal}{\emph{Int. J. Comput. Commun. Control}}
  \bibinfo{volume}{11}, \bibinfo{number}{6} (\bibinfo{year}{2016}),
  \bibinfo{pages}{776--788}.
\newblock


\bibitem[\protect\citeauthoryear{Helsgaun}{Helsgaun}{2000}]%
        {Helsgaun:2000}
\bibfield{author}{\bibinfo{person}{Keld Helsgaun}.}
  \bibinfo{year}{2000}\natexlab{}.
\newblock \showarticletitle{An effective implementation of the
  {Lin–-Kernighan} traveling salesman heuristic}.
\newblock \bibinfo{journal}{\emph{European Journal of Operational Research}}
  \bibinfo{volume}{126}, \bibinfo{number}{1} (\bibinfo{year}{2000}),
  \bibinfo{pages}{106--130}.
\newblock
\showISSN{0377-2217}
\urldef\tempurl%
\url{https://doi.org/10.1016/S0377-2217(99)00284-2}
\showDOI{\tempurl}


\bibitem[\protect\citeauthoryear{Helsgaun}{Helsgaun}{2009}]%
        {Helsgaun:2009}
\bibfield{author}{\bibinfo{person}{Keld Helsgaun}.}
  \bibinfo{year}{2009}\natexlab{}.
\newblock \showarticletitle{General k-opt submoves for the {Lin–-Kernighan}
  {TSP} heuristic}.
\newblock \bibinfo{journal}{\emph{Mathematical Programming Computation}}
  \bibinfo{volume}{1}, \bibinfo{number}{2} (\bibinfo{date}{01 Oct}
  \bibinfo{year}{2009}), \bibinfo{pages}{119--163}.
\newblock
\showISSN{1867-2957}
\urldef\tempurl%
\url{https://doi.org/10.1007/s12532-009-0004-6}
\showDOI{\tempurl}


\bibitem[\protect\citeauthoryear{Johnson, Krishnan, Chhugani, Kumar, and
  Venkatasubramanian}{Johnson et~al\mbox{.}}{2004}]%
        {JohnsonKCKV04}
\bibfield{author}{\bibinfo{person}{David~S. Johnson}, \bibinfo{person}{Shankar
  Krishnan}, \bibinfo{person}{Jatin Chhugani}, \bibinfo{person}{Subodh Kumar},
  {and} \bibinfo{person}{Suresh Venkatasubramanian}.}
  \bibinfo{year}{2004}\natexlab{}.
\newblock \showarticletitle{Compressing Large Boolean Matrices using Reordering
  Techniques.}. In \bibinfo{booktitle}{\emph{VLDB}}. \bibinfo{publisher}{Morgan
  Kaufmann}, \bibinfo{pages}{13--23}.
\newblock


\bibitem[\protect\citeauthoryear{Kieffer and Yang}{Kieffer and Yang}{2000}]%
        {tit/KiefferY00}
\bibfield{author}{\bibinfo{person}{John~C. Kieffer} {and}
  \bibinfo{person}{En{-}Hui Yang}.} \bibinfo{year}{2000}\natexlab{}.
\newblock \showarticletitle{Grammar-based codes: {A} new class of universal
  lossless source codes}.
\newblock \bibinfo{journal}{\emph{{IEEE} Trans. Inf. Theory}}
  \bibinfo{volume}{46}, \bibinfo{number}{3} (\bibinfo{year}{2000}),
  \bibinfo{pages}{737--754}.
\newblock


\bibitem[\protect\citeauthoryear{Kourtis, Goumas, and Koziris}{Kourtis
  et~al\mbox{.}}{2008}]%
        {KourtisGK08}
\bibfield{author}{\bibinfo{person}{Kornilios Kourtis},
  \bibinfo{person}{Georgios~I. Goumas}, {and} \bibinfo{person}{Nectarios
  Koziris}.} \bibinfo{year}{2008}\natexlab{}.
\newblock \showarticletitle{Optimizing sparse matrix-vector multiplication
  using index and value compression}. In \bibinfo{booktitle}{\emph{Conf.
  Computing Frontiers}}. \bibinfo{publisher}{{ACM}}, \bibinfo{pages}{87--96}.
\newblock


\bibitem[\protect\citeauthoryear{Larsson and Moffat}{Larsson and
  Moffat}{2000}]%
        {larsson2000off}
\bibfield{author}{\bibinfo{person}{Jesper Larsson} {and}
  \bibinfo{person}{Alistair Moffat}.} \bibinfo{year}{2000}\natexlab{}.
\newblock \showarticletitle{Off-line dictionary-based compression}.
\newblock \bibinfo{journal}{\emph{Proc. IEEE}} \bibinfo{volume}{88},
  \bibinfo{number}{11} (\bibinfo{year}{2000}), \bibinfo{pages}{1722--1732}.
\newblock


\bibitem[\protect\citeauthoryear{Lohrey}{Lohrey}{2012}]%
        {gcc/Lohrey12}
\bibfield{author}{\bibinfo{person}{Markus Lohrey}.}
  \bibinfo{year}{2012}\natexlab{}.
\newblock \showarticletitle{Algorithmics on SLP-compressed strings: {A}
  survey}.
\newblock \bibinfo{journal}{\emph{Groups Complex. Cryptol.}}
  \bibinfo{volume}{4}, \bibinfo{number}{2} (\bibinfo{year}{2012}),
  \bibinfo{pages}{241--299}.
\newblock


\bibitem[\protect\citeauthoryear{Manning, Raghavan, and Sch{\"{u}}tze}{Manning
  et~al\mbox{.}}{2008}]%
        {MRS:2008}
\bibfield{author}{\bibinfo{person}{Christopher~D. Manning},
  \bibinfo{person}{Prabhakar Raghavan}, {and} \bibinfo{person}{Hinrich
  Sch{\"{u}}tze}.} \bibinfo{year}{2008}\natexlab{}.
\newblock \bibinfo{booktitle}{\emph{Introduction to information retrieval}}.
\newblock \bibinfo{publisher}{Cambridge University Press}.
\newblock


\bibitem[\protect\citeauthoryear{Manzini}{Manzini}{2021}]%
        {KaggleMLM}
\bibfield{author}{\bibinfo{person}{Giovanni Manzini}.}
  \bibinfo{year}{2021}\natexlab{}.
\newblock \bibinfo{title}{A Collection of Some Machine Learning Matrices}.
\newblock
  \bibinfo{howpublished}{\url{https://www.kaggle.com/giovannimanzini/some-machine-learning-matrices}}.
\newblock
\newblock
\shownote{Version 3.}


\bibitem[\protect\citeauthoryear{Mehlhorn and Sanders}{Mehlhorn and
  Sanders}{2008}]%
        {Mehlhorn_Sanders_book_chap11}
\bibfield{author}{\bibinfo{person}{Kurt Mehlhorn} {and} \bibinfo{person}{Peter
  Sanders}.} \bibinfo{year}{2008}\natexlab{}.
\newblock \bibinfo{booktitle}{\emph{Algorithms and Data Structures: The Basic
  Toolbox}}.
\newblock \bibinfo{publisher}{Springer Berlin Heidelberg},
  \bibinfo{address}{Berlin, Heidelberg}, Chapter~11, \bibinfo{pages}{217--232}.
\newblock
\showISBNx{978-3-540-77978-0}
\urldef\tempurl%
\url{https://doi.org/10.1007/978-3-540-77978-0_11}
\showDOI{\tempurl}


\bibitem[\protect\citeauthoryear{Moffat and Petri}{Moffat and Petri}{2020}]%
        {tois/MoffatP20}
\bibfield{author}{\bibinfo{person}{Alistair Moffat} {and}
  \bibinfo{person}{Matthias Petri}.} \bibinfo{year}{2020}\natexlab{}.
\newblock \showarticletitle{Large-Alphabet Semi-Static Entropy Coding Via
  Asymmetric Numeral Systems}.
\newblock \bibinfo{journal}{\emph{{ACM} Trans. Inf. Syst.}}
  \bibinfo{volume}{38}, \bibinfo{number}{4} (\bibinfo{year}{2020}),
  \bibinfo{pages}{33:1--33:33}.
\newblock


\bibitem[\protect\citeauthoryear{Navarro}{Navarro}{2021}]%
        {csur/Navarro21a}
\bibfield{author}{\bibinfo{person}{Gonzalo Navarro}.}
  \bibinfo{year}{2021}\natexlab{}.
\newblock \showarticletitle{Indexing Highly Repetitive String Collections, Part
  {I:} Repetitiveness Measures}.
\newblock \bibinfo{journal}{\emph{{ACM} Comput. Surv.}} \bibinfo{volume}{54},
  \bibinfo{number}{2} (\bibinfo{year}{2021}), \bibinfo{pages}{29:1--29:31}.
\newblock


\bibitem[\protect\citeauthoryear{Ochoa and Navarro}{Ochoa and Navarro}{2019}]%
        {tit/OchoaN19}
\bibfield{author}{\bibinfo{person}{Carlos Ochoa} {and} \bibinfo{person}{Gonzalo
  Navarro}.} \bibinfo{year}{2019}\natexlab{}.
\newblock \showarticletitle{RePair and All Irreducible Grammars are Upper
  Bounded by High-Order Empirical Entropy}.
\newblock \bibinfo{journal}{\emph{{IEEE} Trans. Inf. Theory}}
  \bibinfo{volume}{65}, \bibinfo{number}{5} (\bibinfo{year}{2019}),
  \bibinfo{pages}{3160--3164}.
\newblock


\bibitem[\protect\citeauthoryear{Saad}{Saad}{2003}]%
        {Saad:2003}
\bibfield{author}{\bibinfo{person}{Yousef Saad}.}
  \bibinfo{year}{2003}\natexlab{}.
\newblock \bibinfo{booktitle}{\emph{Iterative methods for sparse linear
  systems}}.
\newblock \bibinfo{publisher}{{SIAM}}.
\newblock


\bibitem[\protect\citeauthoryear{Shi}{Shi}{2020}]%
        {Shi_primary_set}
\bibfield{author}{\bibinfo{person}{Jia Shi}.} \bibinfo{year}{2020}\natexlab{}.
\newblock \showarticletitle{Column Partition and Permutation for Run Length
  Encoding in Columnar Databases}. In \bibinfo{booktitle}{\emph{Proceedings of
  the 2020 ACM SIGMOD International Conference on Management of Data}}
  (Portland, OR, USA) \emph{(\bibinfo{series}{SIGMOD '20})}.
  \bibinfo{publisher}{Association for Computing Machinery},
  \bibinfo{address}{New York, NY, USA}, \bibinfo{pages}{2873–2874}.
\newblock
\showISBNx{9781450367356}
\urldef\tempurl%
\url{https://doi.org/10.1145/3318464.3384413}
\showDOI{\tempurl}


\bibitem[\protect\citeauthoryear{Storer and Szymanski}{Storer and
  Szymanski}{1982}]%
        {jacm/StorerS82}
\bibfield{author}{\bibinfo{person}{James~A. Storer} {and}
  \bibinfo{person}{Thomas~G. Szymanski}.} \bibinfo{year}{1982}\natexlab{}.
\newblock \showarticletitle{Data compression via textual substitution}.
\newblock \bibinfo{journal}{\emph{J. {ACM}}} \bibinfo{volume}{29},
  \bibinfo{number}{4} (\bibinfo{year}{1982}), \bibinfo{pages}{928--951}.
\newblock


\bibitem[\protect\citeauthoryear{The Apache Software Foundation}{The Apache
  Software Foundation}{2021}]%
        {SystemDS}
The Apache Software Foundation \bibinfo{year}{2021}\natexlab{}.
\newblock \bibinfo{title}{{Apache SystemDS}}.
\newblock \bibinfo{howpublished}{\url{https://systemds.apache.org/}}.
\newblock
\newblock
\shownote{[Online; accessed 14-Dec-2021].}


\bibitem[\protect\citeauthoryear{Vo and Vo}{Vo and Vo}{2007}]%
        {VO2007273}
\bibfield{author}{\bibinfo{person}{Binh~Dao Vo} {and}
  \bibinfo{person}{Kiem-Phong Vo}.} \bibinfo{year}{2007}\natexlab{}.
\newblock \showarticletitle{Compressing table data with column dependency}.
\newblock \bibinfo{journal}{\emph{Theoretical Computer Science}}
  \bibinfo{volume}{387}, \bibinfo{number}{3} (\bibinfo{year}{2007}),
  \bibinfo{pages}{273--283}.
\newblock


\end{thebibliography}

\end{document}